\pdfoutput=1
\documentclass[epj]{svjour}
\usepackage{graphics}
\usepackage{amsmath}
\usepackage{amssymb}
\usepackage{tensor}
\usepackage{bm}
\def\Im{\mathop{\rm Im}\nolimits}
\def\sinh{\mathop{\rm sinh}\nolimits}
\def\acosh{\mathop{\rm acosh}\nolimits}
\def\Tr{\mathop{\rm Tr}\nolimits}
\begin{document}
\title{Transport coefficients and resonances for a meson gas in Chiral Perturbation Theory}
\author{D.~Fern\'andez-Fraile\thanks{Electronic address: danfer@fis.ucm.es}
\and A.~G\'omez Nicola\thanks{Electronic address: gomez@fis.ucm.es}
}                     
%
%
\institute{Departamento de F\'{\i}sica Te\'orica II, Universidad Complutense, 28040 Madrid, Spain.}
\date{Received: date / Revised version: date}
%
\abstract{We present recent results on a systematic method to
calculate transport coefficients for a meson gas (in particular, we
analyze a pion gas) at low temperatures in the context of Chiral
Perturbation Theory (ChPT). Our method is based on the study of Feynman
diagrams taking into account collisions in the plasma by means of
the non-zero particle width. This implies a modification of the
standard ChPT power counting scheme. We discuss the importance of
unitarity, which allows for an accurate high energy description of
scattering amplitudes, generating dynamically the $\rho (770)$ and
$f_0(600)$ mesons. Our results are compatible with analyses of
kinetic theory, both in the non-relativistic very low-$T$ regime and
near the transition. We show the behavior with temperature of the
electrical and thermal conductivities as well as of the shear and
bulk viscosities. We obtain that bulk viscosity is  negligible
against shear viscosity, except near the chiral phase transition
where the conformal anomaly might induce larger bulk effects.
Different asymptotic limits for transport coefficients, large-$N_c$
scaling and some applications to heavy-ion collisions are studied.
} 
\titlerunning{Transport coefficients and resonances for a meson gas in ChPT}
\maketitle
\section{Introduction}

The analysis of transport properties within the Heavy-Ion Collision program has become a very interesting topic, with many phenomenological and theoretical implications. Transport coefficients provide the response of the system to thermodynamic forces that take it out of equilibrium. In the linear approximation, energy and momentum transport is encoded in the viscosity coefficients (shear and bulk) whereas charge and heat conduction produce electrical and thermal conductivities, respectively.

A prominent example of physical applications to collisions of heavy ions is found in viscosities. Although the matter produced after thermalization behaves as a nearly  perfect fluid \cite{Adams:2003am}, there are measurable deviations, which are seen mainly in elliptic flow and can be reasonably explained with a small shear viscosity over entropy density ratio \cite{viscotheor}. In these analyses bulk viscosity is customarily neglected, based on several theoretical studies. However, it has been recently noted \cite{Kharzeev,Karsch} that the bulk viscosity might be larger than expected near the QCD phase transition, by the effect of the conformal anomaly. On the other hand, shear viscosity over entropy density is believed to have a minimum in that region. In that case, i.e., if the two viscosity coefficients are comparable at the temperatures of interest, there are several physical consequences such as radial flow suppression, modifications of the hadronization mechanism \cite{Karsch}, or clustering at freeze-out \cite{Torrieri:2007fb}.

Lattice analyses of transport coefficients are cumbersome, since they involve the zero-momentum and energy limit of spectral functions \cite{resco,Meyer,Pica,Gupta}, and these calculations are still not conclusive. It is therefore very interesting and useful to consider regimes accessible to theoretical analysis in order to provide complementary information about transport coefficients. The theoretical approach to transport coefficients has been traditionally carried out within two frameworks: kinetic theory and the diagrammatic approach (Linear Response Theory). The kinetic theory approach involves linearized Boltzmann-like equations and has been successfully applied in high temperature QCD \cite{Arnold} and in the meson sector \cite{Prakash,DobadoFelipe}, while the diagrammatic method has been developed for high-$T$ scalar and gauge theories \cite{Jeon,Valle}. In both formalisms, it is crucial to include accurately the collisional width, identifying the dominant scattering processes in the plasma. In the diagrammatic framework, we have recently studied transport coefficients within Chiral Perturbation Theory \cite{FernandezFraile:05,ffgnepja,Florianopolis,ffgn08}, pertinent for describing the meson sector at low energies and temperatures below the chiral phase transition \cite{Gerber}. Our analysis shows that in order to include properly the effects of the thermal width, the standard rules of Chiral Perturbation Theory have to be modified for this type of calculation.

In this work, we will present a detailed update of our formalism and main results, paying a more detailed attention to several aspects of formal and phenomenological interest. In particular, we provide a thorough derivation of the relevant formulae, emphasizing the link between Kubo's formulae and the definition of transport coefficients through thermodynamic forces and fluxes. Our ChPT diagrammatic formalism is reviewed, describing the different contributions in the modified power counting and discussing specially the role played by the light resonances, which we generate dynamically by unitarizing with the Inverse Amplitude Method. Detailed results for all transport coefficients are given, underlying the connection with existing phenomenological and theoretical calculations, including the predictions of non-relativistic kinetic theory, which we meet in the very low-$T$ regime. We will also discuss the large-$N_c$ behavior, which will provide an interesting check of some of the results obtained.

\section{Chiral Perturbation Theory}
In order to describe the dynamics of the light mesons (pions, kaons and the eta) we will use Chiral Perturbation Theory, which is an effective field theory of QCD for the low-energy regime \cite{Scherer}. It is based on the spontaneous symmetry breaking of the chiral symmetry of the QCD lagrangian (for massless quarks):
\begin{align}
\chi\equiv\mathrm{SU(3)_L\times SU(3)_R\equiv SU(3)_V\times SU(3)_A\longrightarrow SU(3)_V}\ ,\notag
\end{align}
where the axial generators of the chiral algebra are broken leaving only the vector ones. As the result of this symmetry breaking there must appear a number of massless Goldstone bosons equal to the number of broken generators, and they are physically identified with the pions, kaons and eta. In order to construct an effective lagrangian it is necessary to obtain the transformation rules of the Goldstone bosons, $\phi_a$, under the original chiral group. It can be shown \cite{Scherer} that they transform non-linearly, so if we use the exponential parameterization for the Goldstone bosons we have:
\begin{align}
& U(x)\overset{\chi}{\mapsto} R U(x)L^\dagger\ ,\quad \text{with}\ U(x)\equiv\exp\left(\mathrm{i}\frac{\phi(x)}{F_0}\right)\ ,\notag\\
& \text{and}\ \phi(x)\equiv\sum\limits_a\phi_a(x)\lambda_a\ ,
\end{align}
where $R\in SU(N_f)$, $L\in SU(N_f)$, $\lambda_a$ are proportional to the broken generators, and $F_0$ coincides with the Goldstone boson decay constant to lowest order in the chiral expansion (see below and \cite{Scherer}). This non-linear transformation on $U(x)$ implies the following transformations for the Goldstone bosons separately under the vector and axial charges:
\begin{align}
[Q_a^V,\phi_b]=\mathrm{i}f_{abc}\phi_c\ ,\quad [Q_a^A,\phi_b]=g_{ab}(\bm{\phi})\ ,
\end{align}
where $g_{ab}(\bm{\phi})$ is some non-linear function. So we see
that under the unbroken group the Goldstone bosons transform
linearly but they do it non-linearly under the axial charges
corresponding to the broken generators. Once one knows the
transformation rules for the Goldstone bosons it is possible to
construct a lagrangian which describes their dynamics as the most
general expansion in terms of derivatives of the $U(x)$ field that
respects all the symmetries of QCD:
\begin{align}
\mathcal{L}_\mathrm{ChPT}=\mathcal{L}_2+\mathcal{L}_4+\mathcal{L}_6+\ldots\ ,
\end{align}
where the subindex indicates the number of derivatives of the
field $U(x)$. In practice, we will deal only with $\mathcal{L}_2$ and $\mathcal{L}_4$, given explicitly by the expressions \cite{Gerber} (for the $N_f=2$ case):
\begin{align}
\mathcal{L}_2=\frac{F_0^2}{4}\Tr\{\partial_\mu U^\dagger\partial^\mu U+M_0^2(U^\dagger+U)\}\ ,
\end{align}
and
\begin{align}\label{lagrangianL4}
\mathcal{L}_4=&\ \frac{1}{4}l_1(\Tr\{\partial_\mu U^\dagger \partial^\mu U\})^2\notag\\
&+\frac{1}{4}l_2\Tr\{\partial_\mu U^\dagger \partial_\nu U\}\Tr\{\partial^\mu U^\dagger \partial^\nu U\}\notag\\
&-\frac{1}{8}l_4M_0^2\Tr\{\partial_\mu U^\dagger \partial^\mu U\}\Tr\{U+U^\dagger\}\notag\\
&+\frac{1}{16}(l_3+l_4)M_0^4(\Tr\{U^\dagger+U\})^2+h_1 M_0^4\ ,
\end{align}
where $M_0^2\equiv 2B_0 m$ ($m\equiv m_\mathrm{u}\simeq m_\mathrm{d}$ is the quark mass) coincides with the mass of the pion squared to lowest order. The coupling constants $F_0$, $B_0$, $l_i$ and $h_i$ are called the \emph{low-energy constants}, and are energy- and temperature-independent by construction.

In order to deal with this infinite lagrangian we need a way of estimating the contribution from each Feynman diagram of interest, because we do not have an explicit coupling constant. Given a particular scattering amplitude $\mathcal{M}(m_q,p_i)$, where $m_q$ is the mass of the quarks (we will consider $m\equiv m_\mathrm{u}=m_\mathrm{d}$) and $p_i$ the meson external momenta, the dimension $D$ of the diagram is defined by rescaling these parameters in the following way:
\begin{align}
\mathcal{M}(t p_i,t^2 m_q)\equiv t^D\mathcal{M}(p_i,m_q)\ .
\end{align}
Then, the dimension of a particular diagram can be easily computed (Weinberg's Theorem):
\begin{align}\label{WeinbergTheorem}
D=2+\sum\limits_n N_n(n-2)+2L\ ,
\end{align}
where $L$ is the number of loops in the diagram, and $N_n$ the number of vertices coming from the lagrangian $\mathcal{L}_n$. This dimension so defined actually tells us that the contribution from a given diagram is $\mathcal{O}((p/\mathnormal{\Lambda})^D)$, where $p$ represents an energy, momentum, meson mass or temperature. The scale $\mathnormal{\Lambda}$ will be of order $\mathnormal{\Lambda}_E\sim 4\pi F_\pi\simeq 1.2\ \mathrm{GeV}$ for energies, momenta or meson masses\footnote{It can be estimated as the momentum at which the contribution from a one-loop diagram with vertices of $\mathcal{L}_2$ that contributes to meson-meson scattering equals the tree level contribution from $\mathcal{L}_4$ \cite{Georgi}. It can also be estimated as the energy of the lowest resonance, $\mathnormal{\Lambda}_E\sim\ 770\ \mathrm{MeV}$, from where ChPT is expected to fail (see Section \ref{resonances}).}, and of order $\mathnormal{\Lambda}_T\sim 300\ \mathrm{MeV}$ for temperatures\footnote{It can be estimated as the temperature corresponding to an average momentum equal to $\mathnormal{\Lambda}_E$, or as the critical temperature corresponding to the chiral phase transition.}. Therefore, the chiral expansion will be more reliable as we go down in energies and temperatures.
\section{Transport coefficients}
\label{sec:transport}

For a system out of equilibrium there exist thermodynamic forces (gradients of the temperature, the hydrodynamical velocity or the particle density) and fluxes, where the latter try to smooth-out the uniformities produced by the former in order to restore the equilibrium state of the system. Transport coefficients are defined as the coefficients for a series expansion of the fluxes in terms of the thermodynamic forces. Here we will deal with the transport coefficients corresponding to a \emph{linear} expansion, specifically the shear and bulk viscosities, as well as the thermal and DC conductivities. According to relativistic fluid mechanics \cite{deGroot}, the energy-momentum tensor (flux of four-momentum) of a fluid can be decomposed into a reversible and an irreversible part:
\begin{equation}
T^{\mu\nu}\equiv T^{\mu\nu}_\mathrm{R}+T^{\mu\nu}_\mathrm{I}\ ,
\end{equation}
where
\begin{align}
&T^{\mu\nu}_\mathrm{R}\equiv\epsilon U^\mu U^\nu-P\mathnormal{\Delta}^{\mu\nu}\ ,\\
&T^{\mu\nu}_\mathrm{I}\equiv\left[(I_q^\mu+h\mathnormal{\Delta}^{\mu\sigma}N_\sigma)U^\nu+(I_q^\nu+h\mathnormal{\Delta}^{\nu\sigma} N_\sigma)U^\mu\right]+\mathnormal{\Pi}^{\mu\nu}\ ,
\end{align}
with $\epsilon$ the energy density, $P$ the hydrostatic pressure, $U^\mu$ the hydrodynamic velocity defined as the time-like four-vector that verifies $U^\mu(x)U_\mu(x)=1$, $\mathnormal{\Delta}^{\mu\nu}(x)\equiv g^{\mu\nu}-U^\mu(x)U^\nu(x)$ is a projector, $I_q^\mu$ is the heat flow (difference between the energy flow and the flow of enthalpy carried by the particles), $h$ is the enthalpy or heat function per particle, given by $h=(\epsilon+P)/n$, $N^\mu$ is the conserved current (in case there is some in the system), and $\mathnormal{\Pi}^{\mu\nu}$ is called the viscous pressure tensor defined as the irreversible part of the pressure tensor:
\begin{align}
P^{\mu\nu}\equiv T^{\sigma\rho}\tensor{\mathnormal{\Delta}}{^\mu_\sigma}\tensor{\mathnormal{\Delta}}{^\nu_\rho}\equiv- P\mathnormal{\Delta}^{\mu\nu}+\mathnormal{\Pi}^{\mu\nu}\ .
\end{align}
The flows $I_q^\mu$ and $\mathnormal{\Pi}^{\mu\nu}$ can be written as:
\begin{align}
& I_q^\mu\equiv(U_\nu T^{\nu\sigma}-hN^\sigma)\tensor{\mathnormal{\Delta}}{^\mu_\sigma}\ ,\\
& \mathnormal{\Pi}^{\mu\nu}\equiv\bar{\mathnormal{\Pi}}^{\mu\nu}-\mathnormal{\Pi}\mathnormal{\Delta}^{\mu\nu}\ ,
\end{align}
where we have split $\mathnormal{\Pi}^{\mu\nu}$ into a traceless part, $\bar{\mathnormal{\Pi}}^{\mu\nu}$, and a remainder ($\mathnormal{\Pi}=-\tensor{\Pi}{^\mu_\mu}/3$). If we now define the thermodynamical forces as:
\begin{align}
& X\equiv-\nabla^\mu U_\mu\ ,\\
& X_q^\mu\equiv \frac{\nabla^\mu T}{T}-\frac{\nabla^\mu P}{h n}\ ,\\
&\bar{X}^{\mu\nu}\equiv\left[\frac{1}{2}(\tensor{\mathnormal{\Delta}}{^\mu_\sigma}\tensor{\mathnormal{\Delta}}{^\nu_\tau}+\tensor{\mathnormal{\Delta}}{^\nu_\sigma}\tensor{\mathnormal{\Delta}}{^\mu_\tau})-\frac{1}{3}\tensor{\mathnormal{\Delta}}{^{\mu\nu}}\tensor{\mathnormal{\Delta}}{_{\sigma\tau}}\right]\nabla^\sigma U^\tau\ ,
\end{align}
with $T$ being the temperature and $\nabla^\mu\equiv\mathnormal{\Delta}^{\mu\nu}\partial_\nu$, then the \emph{shear viscosity}, $\eta$, \emph{bulk viscosity}, $\zeta$, and \emph{thermal conductivity}, $\kappa$, are defined through the relations:
\begin{align}
& \mathnormal{\Pi}=\zeta X\ ,\\
& I^\mu_q=\kappa T\mathnormal{\Delta}^{\mu\nu}X_\nu^q\ ,\\
& \bar{\mathnormal{\Pi}}^{\mu\nu}=2\eta \bar{X}^{\mu\nu}\ .
\end{align}
At this point it is convenient to pass to the local rest frame (LRF) and recover the more familiar expressions for transport coefficients. In the LRF (we will denote by a tilde the quantities evaluated in this particular frame of reference), $\tilde{U}^\mu=(1,\bm{0})$, $\tilde{\mathnormal{\Pi}}^{00}=0$ (from the definition), and then we have:
\begin{align}\label{viscotensor}
\tilde{T}_{ij}=P\delta_{ij}+\eta\left(\partial_i \tilde{U}_j+\partial_j \tilde{U}_i+\frac{2}{3}\delta_{ij}\partial_k \tilde{U}^k\right)-\zeta\delta_{ij}\partial_k \tilde{U}^k\ .
\end{align}
And for the case of the thermal conductivity:
\begin{align}\label{thermalc}
\tilde{T}^{i0}-h\tilde{N}^i=-\kappa T\left(\frac{\partial_i T}{T}-\frac{\partial_i P}{hn}\right)\ .
\end{align}
Now, using the relativistic Gibbs-Duhem relation
\begin{align}\label{GibbsDuhem}
\frac{1}{n}\nabla^\mu P=h\frac{\nabla^\mu T}{T}+T\nabla^\mu\left(\frac{\mu}{T}\right)\ ,
\end{align}
we can rewrite (\ref{thermalc}) as
\begin{align}\label{thermalc2}
\tilde{T}^{i0}-h\tilde{N}^i=\kappa\frac{T^2}{h}\, \partial_i\left(\frac{\mu}{T}\right)\ .
\end{align}
Therefore we see that for a system without any conserved current (besides the energy-momentum tensor) the thermal conductivity is zero \cite{Danielewicz}. It is also convenient to explicitly write the expression of the thermal conductivity for both the Landau's and Eckart's choices of the hydrodynamical velocity \cite{deGroot}:
\begin{align}
& U_\mathrm{(L)}^\mu(x)\equiv\frac{T^{\mu\nu}U_\nu}{\sqrt{U_\nu T^{\nu\sigma}T_{\sigma\rho}U^\rho}}\quad\Rightarrow\quad \tilde{T}_\mathrm{(L)}^{0i}=0\ ,\\
& U_\mathrm{(E)}^\mu(x)\equiv\frac{N^\mu(x)}{\sqrt{N^\nu(x) N_\nu(x)}}\quad\Rightarrow\quad \tilde{N}_\mathrm{(E)}^i=0\ .
\end{align}
Then,
\begin{align}
&\tilde{N}_\mathrm{(L)}^i=-\kappa\frac{T^2}{h^2}\, \partial_i\left(\frac{\mu}{T}\right)\ ,\\
&\tilde{T}_\mathrm{(E)}^{i0}=\kappa\frac{T^2}{h}\, \partial_i\left(\frac{\mu}{T}\right)\ .
\end{align}
In the non-relativistic limit, we can neglect the second term in
the right hand side of (\ref{thermalc}) and we recover, for the
Eckart's choice, the Fourier's law
$T_\mathrm{(E)}^{0i}=-\kappa\partial_i T$. It is important to
remark here that these Landau's and Eckart's conventions apply to
macroscopic averages of the currents $T^{\mu\nu}$ and $N^\mu$ over
a fluid element, and not to the microscopic currents themselves.
The microscopic quantities will be relevant in the next section,
where we obtain the expressions for transport coefficients in
Linear Response Theory.

\vspace{0.3cm}Finally, for the \emph{DC conductivity}, an electric current is induced in the gas by an external electric field which is constant in space and time, $J^i=\tensor{\sigma}{^i_j}E_\mathrm{ext}^j$. In general, the DC conductivity will be a tensor, but we will consider here the isotropic case, so $\sigma_{ij}=\sigma g_{ij}$.

\subsection{Kubo's formulae for transport coefficients}
Let be a system described by a hamiltonian $\hat{H}_0$ (independent of time) to which we add a perturbation $\hat{V}(t)$, such that $\hat{V}(t)=0$ for $t\leq 0$. Linear Response Theory (LRT) consists in taking into account only the linear effects produced by the perturbation on the magnitudes of the system. For this to be a good approximation is necessary that $\hat{V}(t)$ be small in the sense that the eigenvalues $E_\alpha$ of $\hat{H}(t)\equiv\hat{H}_0+\hat{V}(t)$ and the eigenvalues $E^{(0)}_\alpha$ of $\hat{H}_0$ verify $|E_\alpha-E^{(0)}_\alpha|/E^{(0)}_\alpha\ll 1$. Then, it can be shown \cite{LeBellac} that if $\hat{O}(t)$ is a certain operator in the Schr\"odinger's picture, the variation of the mean value of the operator produced when the perturbation is introduced is given (to linear order in the perturbation) by:
\begin{align}\label{response}
&\delta\langle\hat{O}(t)\rangle\equiv\langle\mathnormal{\Psi}(t)|\hat{O}(t)|\mathnormal{\Psi}(t)\rangle-\langle\mathnormal{\Psi}^{(0)}(t)|\hat{O}(t)|\mathnormal{\Psi}^{(0)}(t)\rangle\notag\\
&=-\mathrm{i}\int\limits_0^\infty\mathrm{d}t'\ \langle\mathnormal{\Psi}^{(0)}(0)|\theta(t-t')[\hat{O}_\mathrm{H}(t),\hat{V}_\mathrm{H}(t')]|\mathnormal{\Psi}^{(0)}(0)\rangle\ ,
\end{align}
where $|\mathnormal{\Psi}^{(0)}(0)\rangle$ represents the state of the system at $t=0$, $|\mathnormal{\Psi}^{(0)}(t)\rangle\equiv\mathrm{e}^{-\mathrm{i}\hat{H}_0 t}|\mathnormal{\Psi}^{(0)}(0)\rangle$, $|\mathnormal{\Psi}(t)\rangle\equiv\mathrm{e}^{-\mathrm{i}\hat{H} t}|\mathnormal{\Psi}^{(0)}(0)\rangle$, and $\hat{O}_\mathrm{H}(t)\equiv\mathrm{e}^{\mathrm{i}\hat{H}_0t}\hat{O}(t)\mathrm{e}^{-\mathrm{i}\hat{H}_0t}$ (Heisenberg's picture). The result (\ref{response}) is also valid if instead of mean values we deal with thermal averages $\langle\ \cdot\ \rangle_T$ and then we want to study small deviations from thermal equilibrium. Note that in this case, according to (\ref{response}), we calculate these deviations by evaluating the expectation value of the commutator at equilibrium.

\vspace{0.3cm}Now, by applying some small perturbation to our system in order to take it slightly out of equilibrium and using LRT, we can obtain the expressions for transport coefficients in terms of correlators. We start considering the DC conductivity. We perturb the system by coupling it to an external classical electromagnetic field:
\begin{align}
\hat{V}(t)=\int\mathrm{d}^3\bm{x}\ \hat{J}_\mu(x) A_\mathrm{ext}^{\mu}(x)\ .
\end{align}
Therefore, the induced current is
\begin{align}
\delta\langle\hat{J}_\mu(x)\rangle=-\mathrm{i}\int\mathrm{d}^4x'\ G^\mathrm{R}_{\mu\nu}(x-x')A^\nu_\mathrm{ext}(x')\ ,
\end{align}
where $G^\mathrm{R}_{\mu\nu}(x)\equiv\theta(t)\langle[\hat{J}_\mu(x),\hat{J}_\nu(0)]\rangle$ is the \emph{retarded} propagator. Now, using the gauge $A_\mathrm{ext}^0=0$, we get $E_\mathrm{ext}^i=-\partial_0A^i_\mathrm{ext}$. And therefore, in momentum space,
\begin{align}
J^i(\omega,\bm{p})=-\frac{\tensor{(G^\mathrm{R})}{^i_j}}{\omega}E^j_\mathrm{ext}(\omega,\bm{p})\equiv\tensor{\sigma}{^i_j}(\omega,\bm{p})E^j_\mathrm{ext}(\omega,\bm{p})\ .
\end{align}
Since the DC conductivity corresponds to the action of a \emph{constant} electric field, and assuming spatial isotropy, we finally have:
\begin{align}\label{DCcond}
\sigma &=-\frac{1}{3}\lim_{\omega\rightarrow 0^+}\lim_{|\vec{p}|\rightarrow 0^+}\frac{\Im\mathrm{i}\tensor{(G^\mathrm{R})}{^i_i}(\omega,\vec{p})}{\omega}\notag\\
&=-\frac{1}{6}\lim_{\omega\rightarrow 0^+}\lim_{|\vec{p}|\rightarrow 0^+}\frac{\rho_\sigma(\omega,|\vec{p}|)}{\omega}\ ,
\end{align}
where $\rho_\sigma=2\Im\mathrm{i}\tensor{(G^\mathrm{R})}{^i_i}$ is the spectral function of the current-current correlator. The order in which the limit is taken is important, since the opposite would correspond to a field constant in time and slightly not constant in space, what would produce a rearrangement of the static charges giving a vanishing electric current. This is called a Kubo formula for the DC conductivity, and we can express it in another useful form in terms of the Wightman function $G^>$ by using the KMS relation $G^<(p)=\mathrm{e}^{-\beta p^0}G^>(p)$ and $\rho=G^>-G^<$ \cite{Jeon,LeBellac}:
\begin{align}
\sigma&=-\frac{1}{6}\lim\limits_{\omega\rightarrow 0^+}\lim\limits_{|\bm{p}|\rightarrow 0^+}\frac{\partial}{\partial\omega}\int\mathrm{d}^4x\ \mathrm{e}^{\mathrm{i}p\cdot x}\langle[\hat{J}^i(x),\hat{J}_i(0)]\rangle\\
&=-\frac{\beta}{3}\lim\limits_{\omega\rightarrow 0^+}\lim\limits_{|\bm{p}|\rightarrow 0^+}\int\mathrm{d}^4x\ \mathrm{e}^{\mathrm{i}p\cdot x}\ \langle\hat{J}^i(x)\hat{J}_i(0)\rangle\ ,
\end{align}
where $\beta\equiv 1/T$, and we implicitly assume thermal
averages. As we show below,  in the perturbative evaluation of
transport coefficients the spatial momentum can be taken equal to
zero from the beginning.

\vspace{0.3cm}Turning to viscosities, as we have seen, they are related to gradients of the hydrodynamical velocity in the fluid. Since we will evaluate the correlators at thermal equilibrium, we can choose a global reference frame that is at rest with the fluid. We will give here a simple derivation of the Kubo formulae for the viscosities and the thermal conductivity (for a more rigorous discussion see \cite{Zubarev}). By performing a boost that depends on the point, we can simulate gradients of the velocity, so the fluid velocity around some point $x_0$ becomes $U^i(x)\simeq x^j\partial_j U^i(x_0)$. Then, this boost implies the change in the energy density $\delta\mathcal{H}=-\bm{U}\cdot\bm{p}$ where $\bm{p}$ is the density of momentum in the fluid. Therefore, this corresponds to the perturbation in the hamiltonian density:
\begin{align}
\mathcal{V}(x)&=-T^{0i}x^j\partial_j U^i\ .
\end{align}
Under this perturbation, the variation in the expectation value of the energy-momentum tensor is
\begin{align}
\delta\langle\hat{T}^{ij}\rangle=\mathrm{i}\int\mathrm{d}^4x'\ t'\, \theta(t-t')\langle[\hat{T}^{ij}(x),\hat{T}^{kl}(x')]\rangle\partial_k U_l\ ,
\end{align}
where we have integrated by parts and used $\partial_\mu T^{\mu\nu}=0$. If we now particularize for the case $\partial_k U^k=0$, and compare with the expression (\ref{viscotensor}), we then obtain:
\begin{align}
\eta=\frac{\mathrm{i}}{12}\int\mathrm{d}^4x\ t\, \theta(t)\langle[\hat{T}_{ij}(x),\hat{T}^{ij}(0)]\rangle\ .
\end{align}
In momentum space, we can write it as
\begin{align}
\eta=\frac{1}{20}\lim_{\omega\rightarrow 0^+}\lim_{|\vec{p}|\rightarrow 0^+}\frac{\rho_\eta(\omega,|\bm{p}|)}{\omega}\ ,
\end{align}
with
\begin{align}
\rho_\eta(\omega,|\bm{p}|)=\int\mathrm{d}^4x\ \mathrm{e}^{\mathrm{i}p\cdot x}\langle[\hat{\pi}_{ij}(x),\hat{\pi}^{ij}(0)]\rangle\ ,
\end{align}
and $\pi_{ij}\equiv T_{ij}-g_{ij}\tensor{T}{^k_k}/3$. In order to obtain the bulk viscosity we consider instead $\partial_i U_j=(1/3)\delta_{ij}\partial_kU^k$, and we have:
\begin{align}
\delta\mathnormal{\Pi}=\delta\langle\hat{\mathcal{P}}\rangle=\mathrm{i}\int\mathrm{d}^4x'\ t'\, \theta(t-t')\langle[\hat{\mathcal{P}}(x),\hat{\mathcal{P}}(x')]\rangle\partial_k U^k\ ,
\end{align}
with $\mathcal{P}\equiv-\tensor{T}{^k_k}/3$. And comparing with (\ref{viscotensor}), we get:
\begin{align}
\zeta=\mathrm{i}\int\mathrm{d}^4x\ t\, \theta(t)\langle[\hat{\mathcal{P}}(x),\hat{\mathcal{P}}(0)]\rangle\ .
\end{align}
In momentum space we can express it as
\begin{align}
\zeta=\frac{1}{2}\lim_{\omega\rightarrow 0^+}\lim_{|\vec{p}|\rightarrow 0^+}\frac{\rho_\zeta(\omega,|\bm{p}|)}{\omega}\ ,
\end{align}
with
\begin{align}
\rho_\zeta(\omega,|\bm{p}|)=\int\mathrm{d}^4x\ \mathrm{e}^{\mathrm{i}p\cdot x}\langle[\hat{\mathcal{P}}(x),\hat{\mathcal{P}}(0)]\rangle\ .
\end{align}

\vspace{0.3cm}We now derive the Kubo expression for the thermal conductivity. As we have seen, it is necessary to have some conserved current in the system (besides $T^{\mu\nu}$) for the thermal conductivity to be non-zero. According to Eq. (\ref{thermalc2}), heat conduction can be produced by a gradient in the chemical potential. In order to create such a gradient, we couple an external gauge field $A_\mathrm{ext}^\mu$ to the conserved current $N^\mu$ ($A_\mathrm{ext}^0$ plays the role of an effective chemical potential), so the perturbation in the hamiltonian density is:
\begin{align}
\mathcal{V}(x)=N_\mu A_\mathrm{ext}^\mu\ .
\end{align}
By choosing $A_\mathrm{ext}^i=0$, integrating by parts, and using $\partial_\mu N^\mu=0$ we obtain:
\begin{align}
\delta I_q^i&=\delta\langle\hat{T}^{i0}-h \hat{N}^i\rangle(x)\\
&=\mathrm{i}\int\mathrm{d}^4x'\ t'\, \theta(t-t')\langle[\hat{\mathcal{T}}^i(x),N^j(x')]\rangle\partial_j A^0_\mathrm{ext}(x')\ ,
\end{align}
with $\mathcal{T}^i\equiv T^{0i}-h N^i$. Thus, by comparing with eq. (\ref{thermalc2}), taking $\partial_j A^0_\mathrm{ext}$ constant, and assuming spatial isotropy, we have that the Kubo formula for $\kappa$ is:
\begin{align}
\kappa=-\frac{\beta}{6}\lim\limits_{\omega\rightarrow 0^+}\lim\limits_{|\bm{p}|\rightarrow 0^+}\frac{\rho_\kappa(\omega,|\bm{p}|)}{\omega}\ ,
\end{align}
with
\begin{align}
\rho_\kappa(\omega,|\bm{p}|)=\int\mathrm{d}^4x\ \mathrm{e}^{\mathrm{i}p\cdot x}\langle[\hat{\mathcal{T}}^i(x),\hat{\mathcal{T}}_i(0)]\rangle\ .
\end{align}
We here have used the following property of Wightman functions involving a conserved current $N^\mu$ (so $\partial_\mu N^\mu=0$): for any operator $\hat{\mathcal{O}}(x)$, and frequency $\omega\neq 0$, it is verified
\begin{align}\label{conservzero}
\int\mathrm{d}^4x\ \mathrm{e}^{\mathrm{i}\omega t}\langle\hat{N}^0(x)\hat{\mathcal{O}}(0)\rangle=0\ .
\end{align}
From the expression for $\rho_\kappa$ we explicitly see that if there is no conserved current in the system (besides $T^{\mu\nu}$), then the thermal conductivity is zero.

\vspace{0.3cm}In other more rigorous derivations of the Kubo formulas \cite{Zubarev,Hosoya,Horsley}, where they do not use energy-momentum conservation in the correlators, the expression of the bulk viscosity involves the operator
\begin{equation}\label{Pmod}
\hat{\mathcal{P}}\equiv-\tensor{\hat{T}}{^k_k}/3-c_s^2\hat{T}^{00}-\mu \hat{N}^0\ ,
\end{equation}
where $c_s$ is the speed of sound in the plasma and $\mu$ the chemical potential. Because of the property (\ref{conservzero}), this operator will give the same result for the bulk viscosity if it is calculated exactly (non-perturbatively). However, in our calculations we will use (\ref{Pmod}) and $\mathcal{T}^i\equiv T^{0i}-h N^i$ (instead of only $-h N^i$) to obtain the bulk viscosity and the thermal conductivity, because we are interested in a \emph{perturbative} calculation using propagators with a non-zero width, so the conservation of the energy-momentum tensor in correlators will not be exact within the level of approximation we will use. As we will see, the extra $c_s^2$ term in (\ref{Pmod}) will be relevant in our approach.
\section{Transport coefficients in high-temperature quantum field theory}

\begin{figure}[h!]
\centerline{\resizebox{0.30\textwidth}{!}{\includegraphics{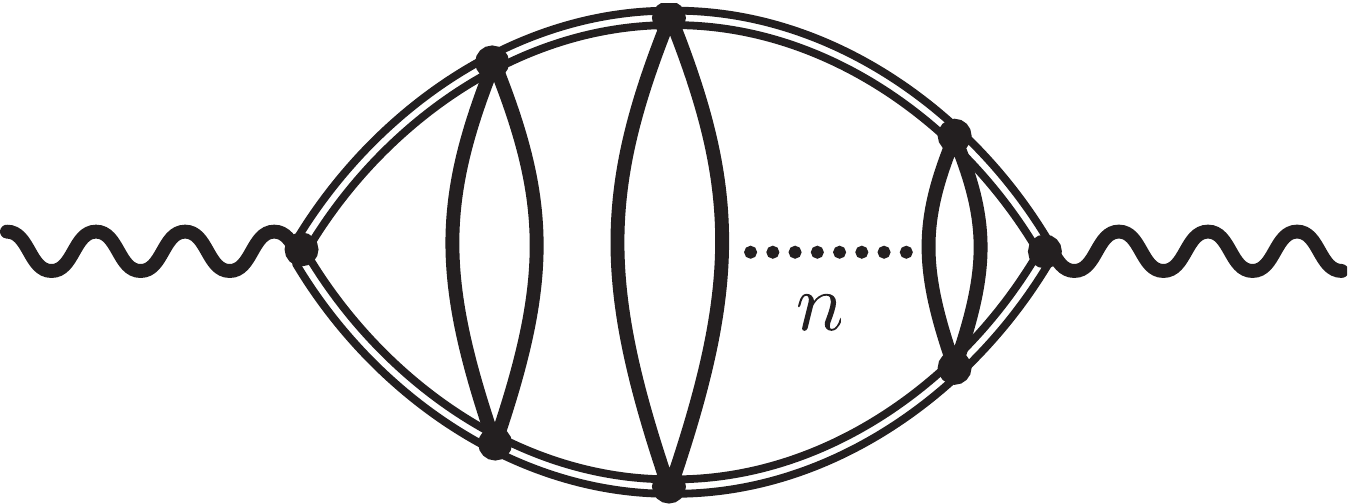}}}
\caption{A ladder diagram with $n$ rungs.} \label{genladderdiag}
\end{figure}

At this point it is convenient to review what happens when one
calculates transport coefficients in high-temperature field
theories \cite{Arnold,Jeon}. In these theories it turns out that
in order to obtain the leading-order result for transport
coefficients a resummation of diagrams is necessary. As we have
seen, a transport coefficient is given in LRT by taking the limit
when the external momentum goes to zero of the imaginary part
(spectral density) of some correlator. This process of taking the
limit of zero external momentum implies the appearance of the so
called \emph{pinching poles}, which are products of retarded and
advanced propagators sharing the same four-momentum:
\begin{align}
G_\mathrm{R}(p)G_\mathrm{A}(p)\simeq\frac{\pi}{4
E_{p}^2\mathnormal{\Gamma}_{p}}\left[\delta(p^0-E_{p})+\delta(p^0+E_{p})\right]\
,
\end{align}
where $\mathnormal{\Gamma}_{p}$ is the particle width (inverse of
the collision time in the plasma) and $E_p$ the particle energy. A
pinching pole would correspond to the contribution from two lines in a
diagram which share the same four-momentum when the external
frequency is zero. For a $\lambda\phi^4$ theory at high temperature,
$\mathnormal{\Gamma}\sim\mathcal{O}(\lambda^2 T)$, so \emph{ladder
diagrams} as the one depicted in Figure \ref{genladderdiag} all count the
same, $\mathcal{O}(1/\lambda^2)$, in the coupling constant and
have to be resummed.

\begin{figure}[h!]
\centerline{\resizebox{0.30\textwidth}{!}{\includegraphics{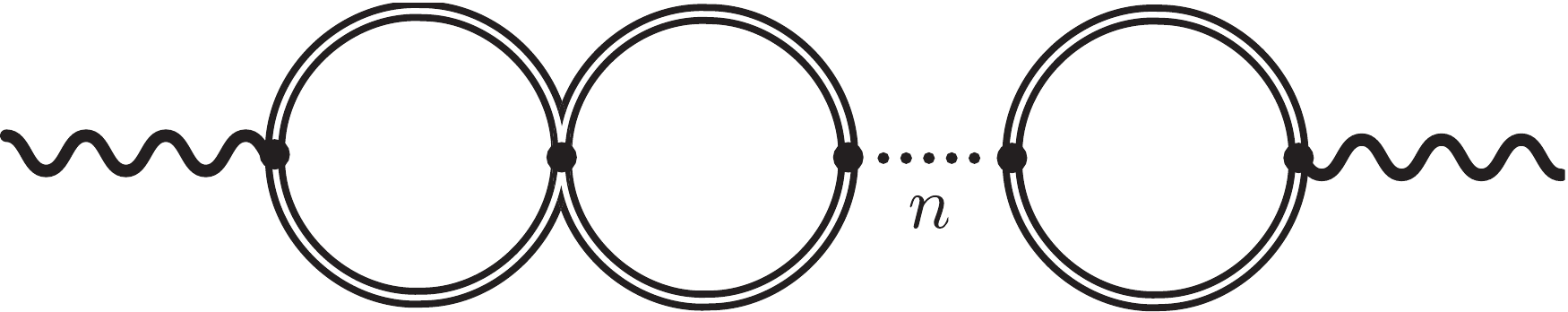}}}
\caption{A bubble diagram with $n$ bubbles.} \label{genbubblediag}
\end{figure}

\begin{figure}[h!]
\centerline{\resizebox{0.20\textwidth}{!}{\includegraphics{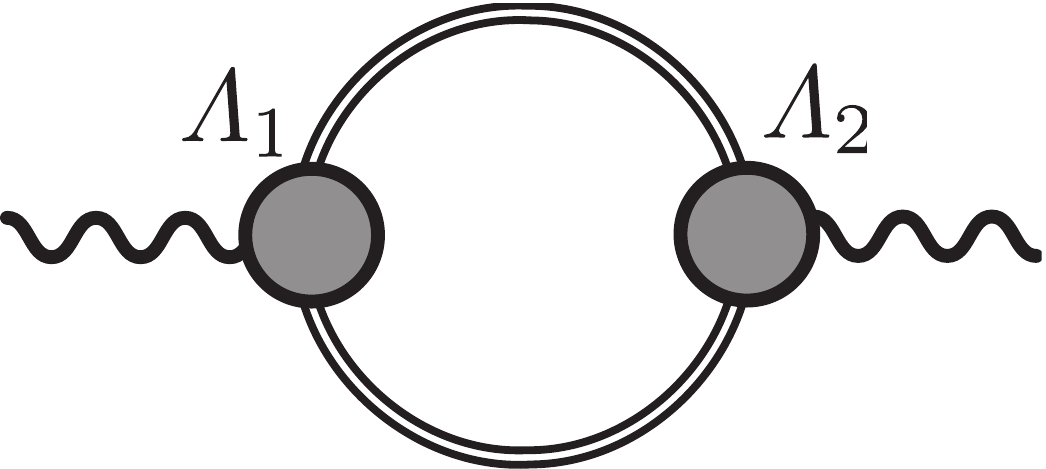}}}
\caption{Generic representation of a ladder and/or bubble diagram in terms of two effective vertices, $\mathnormal{\Lambda}_1$ and $\mathnormal{\Lambda}_2$.} \label{genericdiag}
\end{figure}

Another kind of diagrams, \emph{bubble diagrams} (Fig. \ref{genbubblediag}), in principle would give the dominant contribution, increasing with the number of bubbles according to the counting scheme given above, so they would be naively of order $\mathcal{O}((1/\lambda^2)^n\lambda^{n-1})=\mathcal{O}(1/\lambda^{n+1})$. But after some analysis \cite{Jeon} it can be shown that they can all be resummed giving a subdominant contribution (except for the bulk viscosity) with respect to the one-bubble diagram of Figure \ref{leadingdiag}, i.e., they correspond to the graph of Figure \ref{genericdiag} with $\mathnormal{\Lambda}_1=\mathnormal{\Lambda}^{(0)}+\mathcal{O}(\lambda)$ and $\mathnormal{\Lambda}_2=\mathnormal{\Lambda}^{(0)}$, with $\mathnormal{\Lambda}^{(0)}$ being the lowest-order vertex.

Therefore it is interesting from the theoretical point of view to analyze what happens in ChPT, where we do not have and explicit coupling constant, in order to see whether a resummation is needed.

\section{Particle width}
As we have mentioned in the previous section, it is crucial to have lines dressed in the generic diagram of Fig. \ref{genericdiag} (double lines) in order to take into account the collisions between the particles of the fluid, as dictated also by kinetic theory. If the particle width was zero it would mean that particles would propagate without interaction, implying that the corresponding transport coefficient would be infinite. We can approximate the interaction between the particles in the bath by considering the following spectral density with a non-zero width $\mathnormal{\Gamma}_p$:
\begin{align}
\rho(p^0,\bm{p})=\frac{1}{E_p}\left[\frac{\mathnormal{\Gamma}_p}{(p^0-E_p)^2+\mathnormal{\Gamma}_p^2}-\frac{\mathnormal{\Gamma}_p}{(p^0+E_p)^2+\mathnormal{\Gamma}_p^2}\right] .
\end{align}
This approximation by a Lorentzian will be valid for a small enough width. The width is generically calculated for two-body collisions by \cite{Goity}:
\begin{align}\label{widthGoity}
\mathnormal{\Gamma}(k_1)=\ &\frac{\sinh (\beta E_1/2)}{4E_1}\int\mathrm{d}\nu_2\, \mathrm{d}\nu_3\, \mathrm{d}\nu_4\notag\\
&\times(2\pi)^4\delta^{(4)}(k_1+k_2-k_3-k_4)|T(s,t)|^2\ ,
\end{align}
with
$\displaystyle\mathrm{d}\nu_i\equiv \mathrm{d}^3\bm{k}_i/[(2\pi)^3 2E_i\sinh(\beta E_i/2)]$, $T(s,t)$ is the two-body scattering amplitude, and $s\equiv(k_1+k_2)^2$ and $t\equiv(k_1-k_3)^2$ are the Mandelstam variables.

\begin{figure}[h!]
\centerline{\resizebox{0.20\textwidth}{!}{\includegraphics{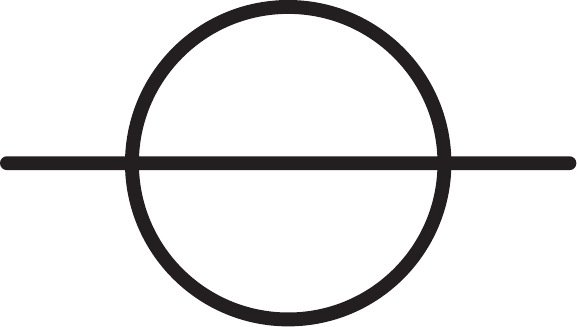}}}
\caption{Leading-order contribution to the pion width.} \label{leadingwidth}
\end{figure}

\vspace{0.3cm}If the gas is dilute, i.e. $\beta E\gg 1$ (dilute gas approximation, DGA), the previous expression reduces to:
\begin{align}
\mathnormal{\Gamma}(k_1)\simeq\frac{1}{2}\int\frac{\mathrm{d}^3
k_2}{(2\pi)^3}\ \mathrm{e}^{-E_2/T}\ \sigma_\text{tot}(s)
v_\text{rel}(1-\bm{v}_1\cdot\bm{v}_2)\ , \label{gammadg}
\end{align}
where $\sigma_\text{tot}$ is the total pion-pion scattering cross
section, $\bm{v}_i$ the velocity of each the two colliding pions,
and $v_\mathrm{rel}$ their relative speed. Up to energies of $1\
\mathrm{GeV}$ it can be shown \cite{AGNJR} that for $\pi\pi$ scattering only the channels
$IJ=00,11,20$ of isospin-angular momentum are relevant and then we can make the approximation:
\begin{align}
\sigma_\text{tot}(s)&=\frac{32\pi}{3s}\sum_{I,J}(2J+1)(2I+1)|t_{IJ}(s)|^2\notag\\
&\simeq\frac{32\pi}{3s}\left[|t_{00}(s)|^2+9 |t_{11}(s)|^2+5 |t_{20}(s)|^2\right]\ ,
\end{align}
where $t_{IJ}(s)$ are the partial waves, so the total scattering
amplitude for $\pi\pi$ scattering is decomposed in terms of the
isospin-projected scattering amplitude, $T_I$, and the partial
waves as:
\begin{align}
& T(s,t)=\frac{1}{3}\sum\limits_{I=0}^2(2I+1)|T_I(s,t)|^2\ ,\\
&\text{with}\quad\displaystyle T_I(s,t)\equiv 32\pi\sum\limits_{J=0}^\infty(2J+1)t_{IJ}(s)P_J(\cos\theta)\ ,
\end{align}
and $P_J$ being Legendre polynomials. Furthermore, in the 00 and 11 channels there appear the $f_0(600)$ and $\rho(770)$ resonances respectively. In order to deal properly with these resonances within ChPT we will have to unitarize our scattering amplitudes (see next section). The leading order contribution to the pion width is represented by the diagram in Figure \ref{leadingwidth}. Cutting this diagram in order to extract its imaginary part (the width) leads to the formula (\ref{widthGoity}) with $T(s,t)$ being the pion-pion scattering amplitude. All the diagramatic calculation will be carried out in the Imaginary Time Formalism (ITF) \cite{LeBellac} which has the advantage of dealing with the same fields, vertices and diagrams as the corresponding vacuum field theory but with properly modified Feynman rules.

\section{Resonances}\label{resonances}
It is difficult for ChPT to deal with the resonances that appear in some of the scattering channels because the unitarity condition
\begin{align}\label{unitcond}
&\hat{S}\hat{S}^\dagger=1\Rightarrow\Im t_{IJ}(s)=\sigma(s)|t_{IJ}(s)|^2\ ,\notag\\
&\text{with}\ \sigma(s)\equiv\sqrt{1-4M_\pi^2/s}\ ,
\end{align}
is not respected for high-enough energy $\sqrt{s}$. This is
because the partial waves calculated in ChPT are essentially
polynomials in $p$ (and logarithms):
$t_{IJ}(s)=t_{IJ}^{(1)}(s)+t_{IJ}^{(2)}(s)+\mathcal{O}(s^3)$, with
$t_{IJ}^{(k)}(s)=\mathcal{O}(s^k)$ and $s=\mathcal{O}(p^2)$. In
order to extend the range of applicability of the next-to-leading
order results for the partial waves calculated in ChPT we will
unitarize them by means of the Inverse Amplitude Method (IAM). The
idea behind this method is essentially to construct an expression
for the scattering amplitude which respects unitarity exactly and
when expanded perturbatively matches to a given order the standard
ChPT expansion. The construction of this amplitude can be
justified more formally by using dispersion relations
\cite{Dobado,Rios}. According to the IAM, the unitarized partial
waves to order $\mathcal{O}(p^4)$ are given by:

\begin{align}
t_{IJ}^\mathrm{U}(s)=\frac{t_{IJ}^{(1)}(s)}{1-t_{IJ}^{(2)}(s)/t_{IJ}^{(1)}(s)}\ .
\end{align}

Using this unitarization method, the $f_0(600)$ and $\rho(770)$ resonances that appear in the pion-pion scattering channels $IJ=00,11$ respectively are correctly reproduced for some set of values of the low-energy constants $\bar{l}_i$ (an overline denotes the renormalized low-energy constant, see (\ref{lagrangianL4})). In addition to appearing as peaks in the scattering cross section, resonances can also be identified as poles in the scattering amplitude after continued to the second Riemann sheet (SRS). If $t^{(\mathrm{I})}$ denotes the analytical continuation of the scattering amplitude off the real axis, then the scattering amplitude on the SRS, $t^{(\mathrm{II})}$, is defined by $\Im t^{(\mathrm{II})}(s-\mathrm{i}0^+)=\Im t^{(\mathrm{I})}(s+\mathrm{i}0^+)$, for $s>4M_\pi^2$. Therefore one has
\begin{align}
t^{(\mathrm{II})}(s)=\frac{t^{(\mathrm{I})}(s)}{1-\mathrm{i}2\sigma(s)t^{(\mathrm{I})}(s)}\ .
\end{align}
A resonance corresponds to a pole of $t^{(\mathrm{II})}$ in the lower half complex plane, being the position of the pole related to the mass and width of the resonance by $s_\mathrm{pole}=(M_R-\mathrm{i}\mathnormal{\Gamma}_R/2)^2$, assuming that the resonance is a narrow Breit-Wigner one, which in the case of the $f_0(600)$ is not a so good approximation. Since we work in the center of mass reference frame, the mass and width obtained correspond to a resonance at rest. In what follows, we will fix the low-energy constants to the values $\bar{l}_1=-0.3$, $\bar{l}_2=5.6$, $\bar{l}_3=3.4$, $\bar{l}_4=4.3$, which imply a mass and width at $T=0$ for the $f_0(600)/\sigma$ resonance of $M_\sigma=441\ \mathrm{MeV}$ and $\mathnormal{\Gamma}_\sigma=464\ \mathrm{MeV}$ respectively, and for the $\rho(770)$ resonance $M_\rho=756\ \mathrm{MeV}$ and $\mathnormal{\Gamma}_\rho=151\ \mathrm{MeV}$.

\begin{figure}[h!]
\centerline{\resizebox{0.40\textwidth}{!}{\includegraphics{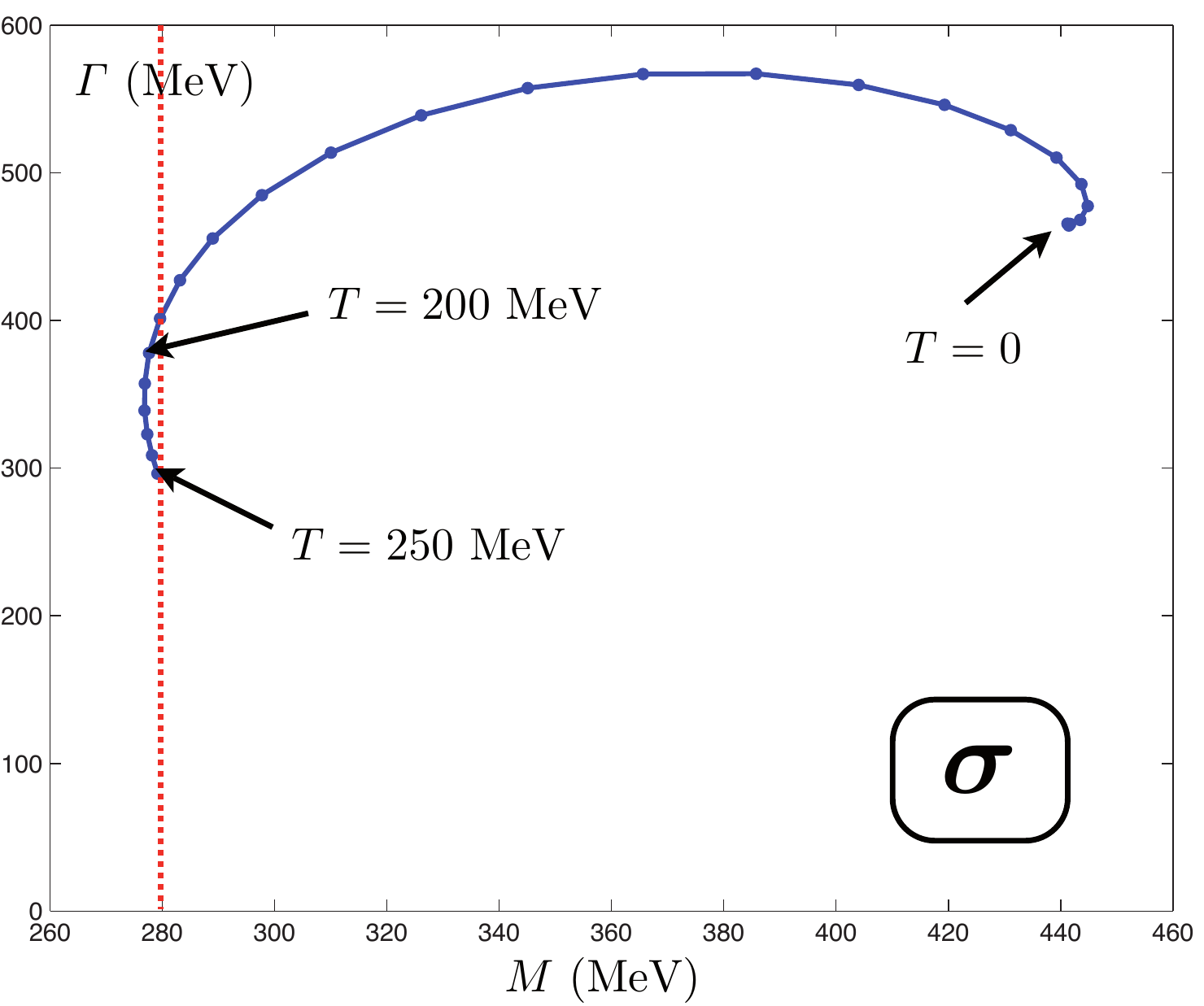}}}
\caption{Evolution of the $f_0(600)/\sigma$ with temperature in the mass-width plane. The red dashed line corresponds to the two-pion threshold. Each dot corresponds to an increment of $10\ \mathrm{MeV}$ in temperature. The $\sigma$ remains broad near the phase transition although it experiences a big decrease in its mass.} \label{sigmapoleT}
\end{figure}

\vspace{0.3cm}If the pion gas is dilute enough so only intermediate two-pion states are relevant in the thermal bath, then we can define \emph{thermal} scattering amplitudes as those calculated like in the $T=0$ case but using thermal propagators instead \cite{GomezNicola:02,FernandezFraile:07}. The thermal scattering amplitudes $t(s,T)$ (the temperature dependence starts to show up at $\mathcal{O}(s^2)$) also must verify the unitarity condition (\ref{unitcond}) but making the replacement $\sigma(s)\mapsto\sigma_T(s,T)\equiv\sigma(s)[1+2n_\mathrm{B}(\sqrt{s}/2)]$ (thermal phase space)\footnote{$n_\mathrm{B}(x)\equiv (\mathrm{e}^{x/T}-1)^{-1}$ is the Bose-Einstein distribution function.} there, and also in the expressions for the unitarized amplitudes. By considering these scattering amplitudes, we can study the evolution with temperature of the poles corresponding to the $f_0(600)$ and $\rho(770)$ resonances. In Fig. \ref{sigmapoleT} we plot the evolution of the $\sigma$ pole with temperature, and we see that it remains broad at temperatures near the phase transition, while its mass is driven to the threshold as a possible indication of chiral symmetry restoration.

\begin{figure}[h!]
\centerline{\resizebox{0.40\textwidth}{!}{\includegraphics{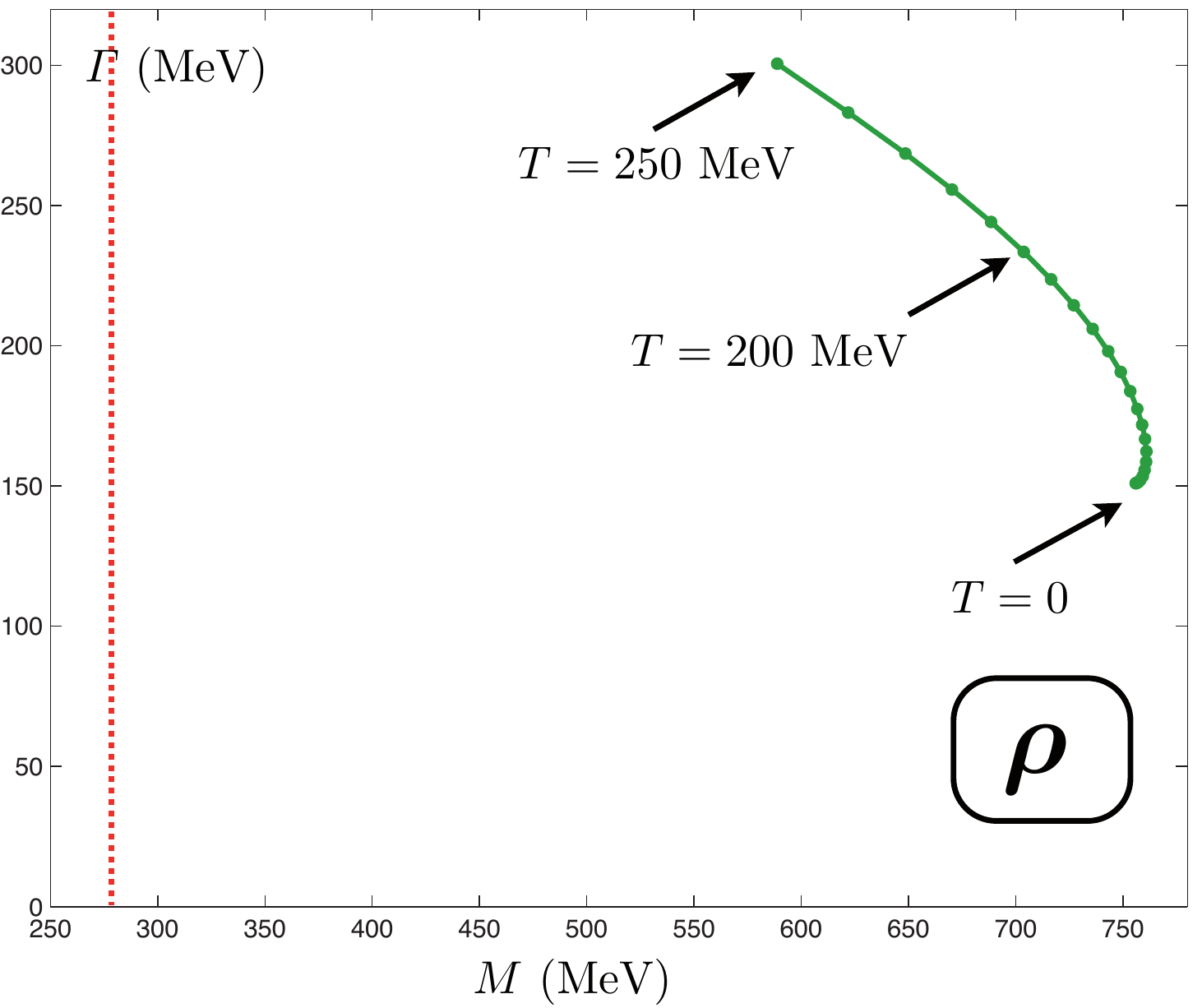}}}
\caption{Evolution of the $\rho(770)$ with temperature in the mass-width plane. The $\rho$ increases its width with temperature and remains far from the threshold at temperatures near the phase transition.} \label{rhopoleT}
\end{figure}

Analogously, in Fig. \ref{rhopoleT} we plot the evolution of the
$\rho$ pole with temperature. We see that the mass shift is small
remaining far from the threshold near the temperature of the chiral
phase transition, and the width increases instead. This behavior is
compatible with the dilepton data from heavy ion collisions (see
details in \cite{FernandezFraile:07,Cabrera}).

\begin{figure}[h!]
\centerline{\resizebox{0.40\textwidth}{!}{\includegraphics{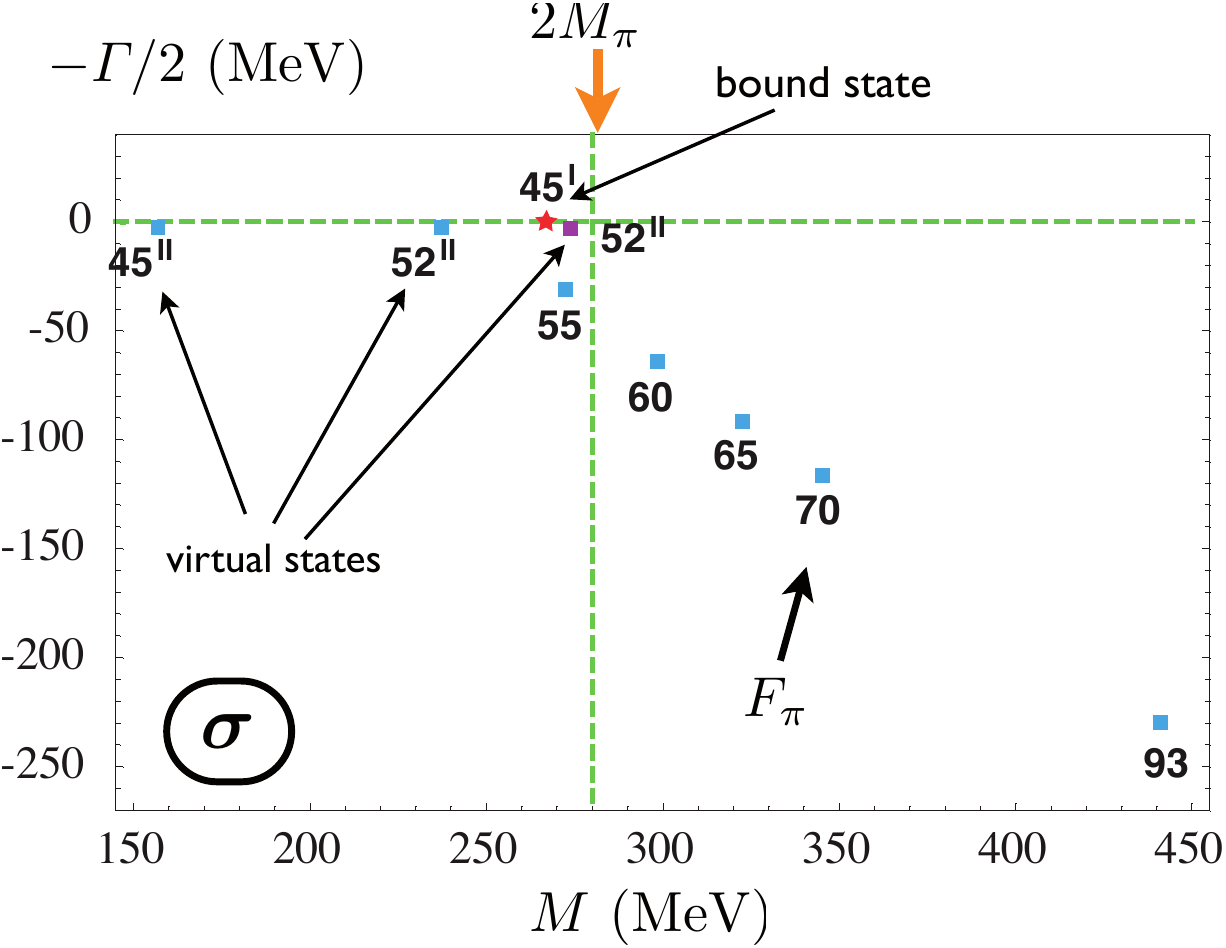}}}
\caption{Evolution of the $f_0(600)/\sigma$ pole at $T=0$ for several nuclear densities controlled effectively by decreasing the pion decay constant. The corresponding values for $F_\pi$ are indicated besides each pole. When the pole crosses the threshold it splits into two virtual states, and for higher densities one becomes a bound state.} \label{sigmapoleFPI}
\end{figure}

We can also study the qualitative evolution of the resonance poles in presence of nuclear density, introduced effectively by varying the pion decay constant $F_\pi$, because
\begin{align}
\frac{F_\pi^2(\rho)}{F_\pi^2(0)}\simeq\frac{\langle\bar{q}q\rangle(\rho)}{\langle\bar{q}q\rangle(0)}\simeq\left(1-\frac{\sigma_{\pi N}}{M_\pi^2F_\pi^2(0)}\rho\right)\simeq\left(1-0.35\frac{\rho}{\rho_0}\right)\ ,
\end{align}
where $\rho$ is the nuclear density, $\sigma_{\pi N}\simeq 45\ \mathrm{MeV}$ is the pion-nucleon sigma term, and $\rho_0\simeq 0.17\ \mathrm{fm}^{-3}$ is the saturation density of nuclear matter. This approach only takes into account a limited class of contributions, but we reproduce several aspects of the expected chiral symmetry restoration behavior at finite density. We have recently \cite{Cabrera} improved the implementation of nuclear density effects for the $f_0(600)$ resonance by considering a microscopic calculation of many-body pion dynamics and unitarizing by solving the Bethe-Salpeter equation, obtaining in this way results qualitatively compatible with this simpler method we analyze here. In Fig. \ref{sigmapoleFPI} we now plot the behavior of the $\sigma$ pole at $T=0$ for several nuclear densities (the corresponding values of $F_\pi$ are indicated besides each pole). We see that density effects drive the $\sigma$ pole faster towards the real axis, becoming a zero-width state, but the required nuclear density is very high (for instance, $F_\pi=55\ \mathrm{MeV}$ is equivalent to $\rho\simeq 1.9\rho_0$). At high enough density, when the pole crosses the threshold, there appear two separated poles on the real axis of the SRS below the threshold (virtual states), and for higher densities one of the two poles becomes a bound state (pole on the real axis in the first Riemann sheet). The virtual state which remains near the threshold and eventually becomes a bound state would correspond to a $\pi\pi$ molecule \cite{Cabrera}, while the other virtual state behaves like the chiral partner of the pion in the sense that it tends to become degenerate in mass with it.

\begin{figure}[h!]
\centerline{\resizebox{0.40\textwidth}{!}{\includegraphics{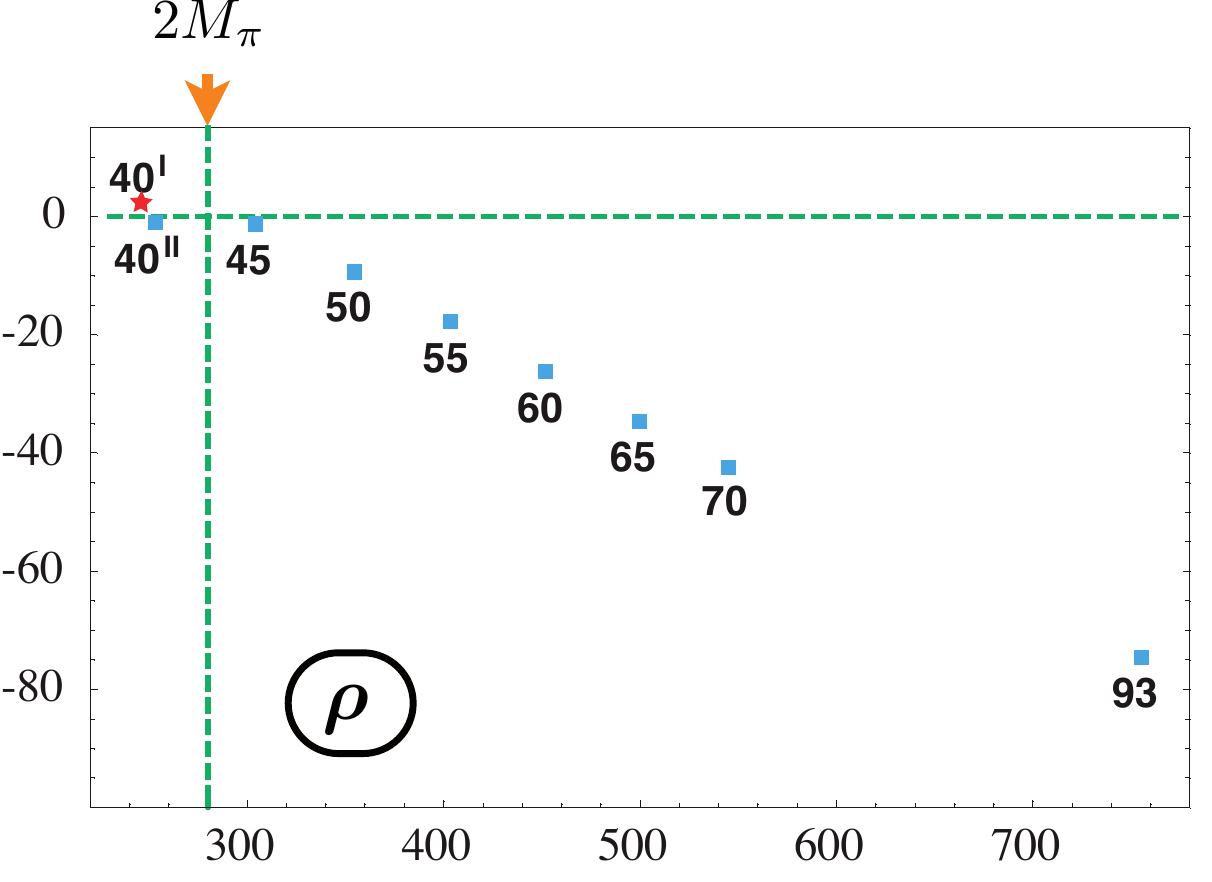}}}
\caption{Evolution of the $\rho(770)$ pole at $T=0$ for several nuclear densities controlled effectively by decreasing the pion decay constant. In this case, after crossing the threshold, the pole becomes a pair virtual state-bound state at almost the same position.} \label{rhopoleFPI}
\end{figure}

Analogously for the $\rho(770)$, we see in Fig. \ref{rhopoleFPI} that density effects also drive it faster toward the real axis and to the threshold. In this case however, after crossing the threshold, the pole becomes a pair virtual state-bound state located at almost the same value of mass and width, indicating a clear $\bar{q}q$ nature for this resonance \cite{Cabrera}.

In the following sections we will study the influence of unitarized
scattering amplitudes on transport coefficients, as well as the
influence of the in-medium evolution of resonances on them.
\section{General analysis of diagrams for transport coefficients in ChPT}
\begin{figure}[h!]
\centerline{\resizebox{0.20\textwidth}{!}{\includegraphics{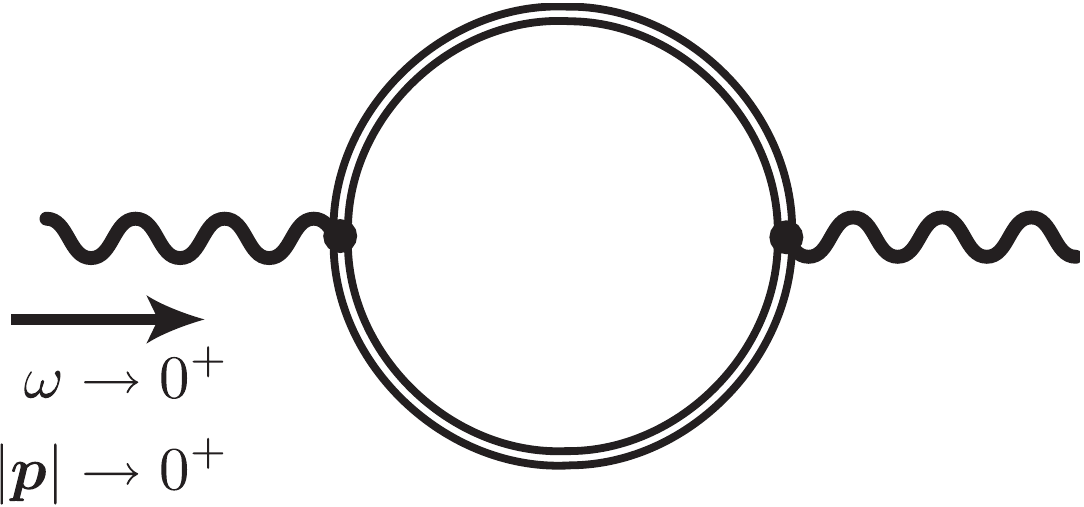}}}
\caption{Leading-order contribution to transport coefficients for low temperatures in ChPT. Double lines represent pion propagators with a non-zero thermal width.} \label{leadingdiag}
\end{figure}

In the analysis of transport coefficients within ChPT, analogously
 to what happens in high-temperature quantum field theory, we
also find non-perturbative contributions, $\propto
1/\mathnormal{\Gamma}$ (and
$\mathnormal{\Gamma}=\mathcal{O}(p^6)$), due to the presence of
pinching poles. This would indicate that the standard ChPT power
counting, dictated by Weinberg's formula (\ref{WeinbergTheorem}),
has to be modified in some way because otherwise, naively,
diagrams with a larger number of pinching poles would become more
important as the temperature is lowered. We will show that for low
temperatures, ladder diagrams are the most relevant, but still
perturbatively small in comparison to the leading order given by
the simple diagram of Fig. \ref{leadingdiag}.

Again, the same topology arguments used in high-temperature
theories are a priori applicable for the ChPT case, so we expect
that the dominant contribution to transport coefficients come from
ladder and bubble diagrams. We start by analyzing the spectral
density corresponding to ladder diagrams. The spectral density of
a generic diagram of the form shown in Fig. \ref{genericdiag} can
be easily calculated in ITF \cite{Valle} obtaining:
\begin{align}\label{formulavalle}
\lim\limits_{\omega\rightarrow 0^+}\frac{\rho(\omega,0)}{\omega}=&\ C\beta\int\frac{\mathrm{d}^3\bm{p}}{(2\pi)^3}\int\limits_{-\infty}^\infty\frac{\mathrm{d}\xi}{2\pi}\ n_B(\xi)[1+n_\mathrm{B}(\xi)]\notag\\
&\times G_\mathrm{R}(\xi,\bm{p})G_\mathrm{A}(\xi,\bm{p})\mathnormal{\Lambda}_1(\xi,\bm{p})\mathnormal{\Lambda}_2(\xi,\bm{p})\ ,
\end{align}
where $C$ is some combinatoric factor which depends on the kind of external field we couple to the pion loop, and it can be shown \cite{FernandezFraile:05} that when considering a non-zero width we can take the zero spatial-momentum limit from the beginning. In the case of the simple diagram without rungs of Fig. \ref{leadingdiag}, and for some constant external insertions, at $T\ll M_\pi$ we obtain that the spectral density behaves like $\lim_{\omega\rightarrow 0^+}\rho(\omega)/\omega\sim\sqrt{M_\pi/T}$, indicating that there could be important non-perturbative contributions from higher-order diagrams (ladder diagrams with an arbitrary number of rungs) in the low-temperature regime. In order to give a first and naive estimation of the contribution at low temperatures ($T\ll M_\pi$) from every diagram we assign a factor $Y$, that we expect to be of order $Y_1\equiv\sqrt{M_\pi/T}$ for very low temperatures, to each pair of lines sharing the same four-momentum, and a factor $X$ that we expect to be of order $X_1\equiv[M_\pi/(4\pi F_\pi)]^2$  for very low temperatures, to any other ``ordinary'' loop ($X_1$ is the typical contribution from a chiral loop). Therefore, according to this new counting, the contribution from a ladder diagram with $n$ rungs would be of order $\mathcal{O}(X^nY^{n+1})$, so ladder diagrams could in principle become more important as we go down in temperatures (where ChPT is expected to work better). Evidently, the contribution from the simple diagram in Fig. \ref{leadingdiag} would be of order $\mathcal{O}(Y)$ instead of the $\mathcal{O}(X)$ estimation given by Weinberg's power counting. In order to verify this naive counting we have explicitly performed \cite{FernandezFraile:05} the resummation of all the ladder diagrams for $T\ll M_\pi$ and have found that it corresponds to multiply the lowest order result from Fig. \ref{leadingdiag} by some constant factor. This is because the contribution $X$ from ordinary loops at very low temperatures is much smaller than naively expected, so $X\sim X_2\equiv\sqrt{T/M_\pi}$ for $T\ll M_\pi$. Therefore the actual contribution from ladder diagrams at very low temperatures is $\mathcal{O}(X_2^n Y_1)$, so they are perturbatively suppressed when the number of rungs increases. However, although ladder diagrams give a contribution much smaller than naively expected, their actual contribution, $\mathcal{O}(p^{2n})$, is still much larger than the estimated by Weinberg's counting, i.e., $\mathcal{O}(p^{4n})$. Since in our ChPT lagrangian we also have derivative vertices, we expect that as the temperature increases, derivative vertices begin to dominate and the loop factor $X$ increases over $X_1$. Also, the pinching pole factor would eventually become of order $\mathcal{O}(1)$ and at temperatures near the phase transition, in principle, we would have to resum all the ladder diagrams.

\vspace{0.3cm}Regarding  bubble diagrams, it can be shown
\cite{FernandezFraile:05} that they all can be resummed giving a
vanishing contribution in the limit $\omega\rightarrow 0^+$, for
the leading order in $1/\mathnormal{\Gamma}$ at very low
temperatures. Other diagrams, like ladder or bubble diagrams with
vertices coming from the lagrangians $\mathcal{L}_n$ with $n>2$,
or diagrams with loops made with more than four-pions vertices
would also be suppressed by the same arguments. But as we have
commented before, as temperature increases $Y$ becomes of order
$\mathcal{O}(1)$ and the particle width is not small anymore, so a
resummation of all the possible diagrams would be in principle
necessary. In this paper however, we will show the results
corresponding to extrapolations of the leading order at low
temperatures, and we will see that these extrapolations, when
unitarized, give the correct order of magnitude for some
observables near the phase transition, which indicates that the
high temperature improvement due to unitarization is a key feature
in this approach.

We also remark that in the limit $T\ll M_\pi$, exponentials like
that  in (\ref{gammadg}) select small three-momenta of
$\mathcal{O}(\sqrt{M_\pi T})$. Therefore, in this limit it is
enough to consider only the $\mathcal{O}(p^2)$ amplitudes in the
cross section, which allows to perform a systematic $T/M_\pi$
expansion \cite{FernandezFraile:05} for $\Gamma_p$ and transport
coefficients, whose leading order results we give below.

In the opposite limit we have massless pions (chiral limit) which
is expected to be reached asymptotically at high temperatures. For
$M_\pi=0$ and using only the $\mathcal{O}(p^2)$ amplitudes, the
thermal width in (\ref{gammadg}) reduces to $\Gamma_p= 5 T^4 |\bm{p}|/(12
\pi^3 F_\pi^4)$ and closed analytic expressions for transport
coefficients can be given. However, these expressions should not
be trusted at temperatures close to the phase transition, where
three-momenta in the integrals are now of $\mathcal{O}(T)$ and
therefore the low-$p$ expansion is not justified and including
unitarity corrections in the amplitude is crucial, as our results
below will show.

\section{Electrical conductivity}
As we have seen, the DC conductivity is given in LRT by:
\begin{align}
\sigma=-\frac{1}{6}\lim\limits_{\omega\rightarrow 0^+}\lim\limits_{|\bm{p}|\rightarrow 0^+}\frac{\rho_\sigma(\omega,|\bm{p}|)}{\omega}\ ,
\end{align}
with
\begin{align}
\rho_\sigma(\omega,|\bm{p}|)=\int\mathrm{d}^4x\ \mathrm{e}^{\mathrm{i}p\cdot x}\langle[\hat{J}^i(x),\hat{J}_i(0)]\rangle\ .
\end{align}
Using the external sources method \cite{Scherer}, we couple an external
electromagnetic field to our ChPT lagrangian and, counting the
electric charge $e=\mathcal{O}(p/\Lambda)$, we calculate the
lowest-order contribution to the DC conductivity, $\sigma^{(0)}$,
given by the diagram of Fig. \ref{leadingdiag}. Then, using the
formula (\ref{formulavalle}) we obtain \cite{FernandezFraile:05}:
\begin{align}
\sigma^{(0)}=\frac{e^2}{3T}\int\frac{\mathrm{d}^3\bm{p}}{(2\pi)^3}\ \frac{|\bm{p}|^2}{E_p^2\mathnormal{\Gamma}_p}n_\mathrm{B}(E_p)[1+n_\mathrm{B}(E_p)]\ ,
\end{align}
where $e$ is the charge of the electron and
$E_p\equiv\sqrt{M_\pi^2+|\bm{p}|^2}$. For very low temperatures,
$T\ll M_\pi$, this expression adopts the simple form:
\begin{align}\label{elcondlowt}
\sigma^{(0)}\simeq 15\frac{e^2 F_\pi^4}{T^{1/2}M_\pi^{5/2}}\ .
\end{align}
It is interesting to compare our result with the expected kinetic theory (KT) behavior. According to KT \cite{Landau}, $\sigma\sim e^2n_\mathrm{ch}\tau/M_\pi$ ($n_\mathrm{ch}$ is the density of charged particles, $\tau$ is the collision mean time, and $e$ is the particle charge), and $\tau\sim1/\mathnormal{\Gamma}$, $\mathnormal{\Gamma}\sim n v\sigma_{\pi\pi}$ ($v$ is the mean speed of the particles). In the non-relativistic limit, $n\sim (\sqrt{M_\pi T})^3\mathrm{e}^{-M_\pi/T}$, $v\sim\sqrt{T/M_\pi}$, and $\sigma_{\pi\pi}$ is a constant, therefore $\sigma\sim 1/\sqrt{T}$. Thus, our result in ChPT is consistent with KT for $T\ll M_\pi$.

\begin{figure}[h!]
\centerline{\resizebox{0.40\textwidth}{!}{\includegraphics{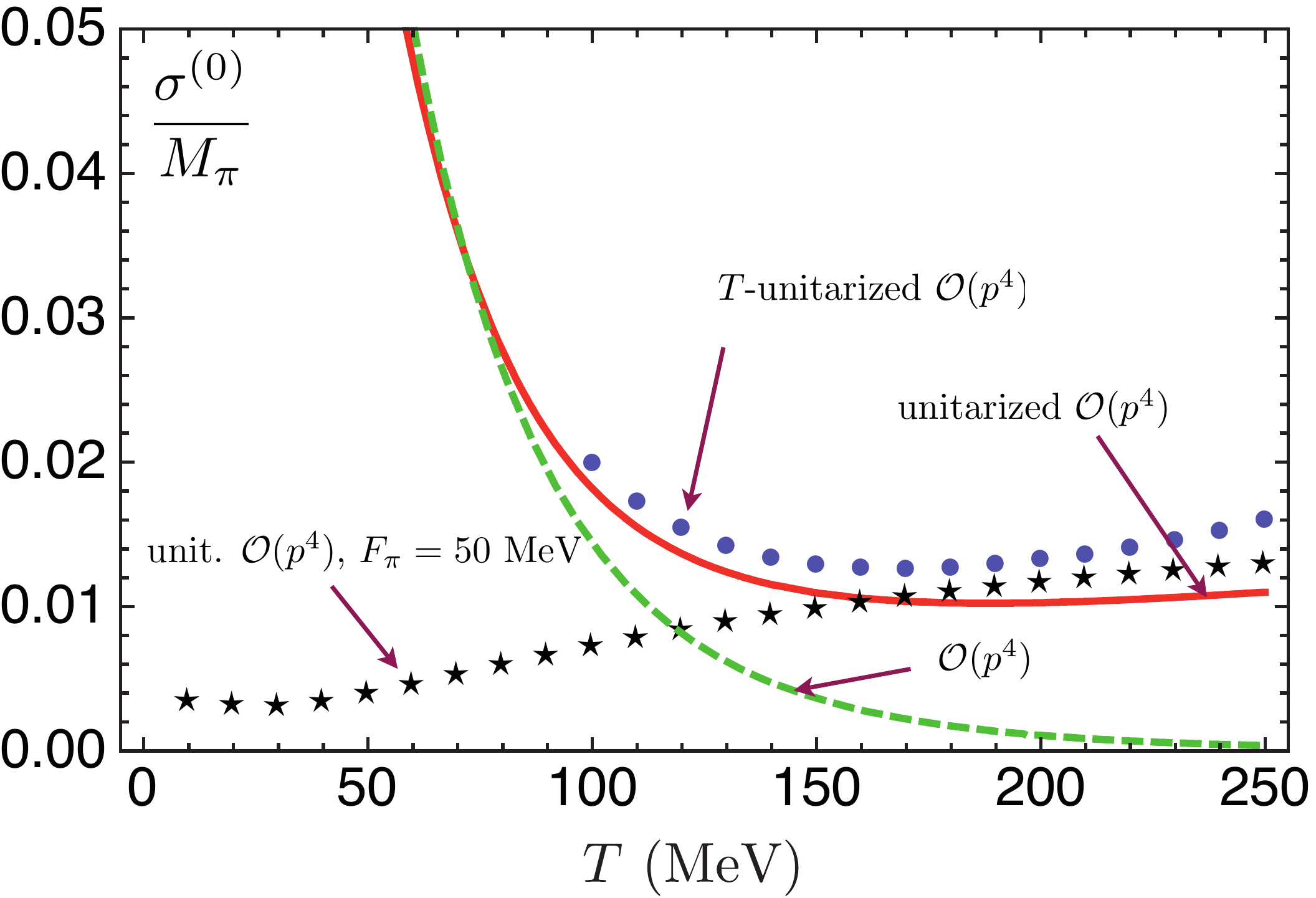}}}
\caption{Lowest-order contribution to the DC conductivity as a function of the temperature. The dashed line corresponds to considering non-unitarized partial waves to $\mathcal{O}(p^4)$ in the pion width, the red line corresponds to the unitarized case to $\mathcal{O}(p^4)$, and the dots correspond to unitarize at $\mathcal{O}(p^4)$ at finite temperature. The stars correspond to unitarized partial waves to $\mathcal{O}(p^4)$ with $F_\pi=50\ \mathrm{MeV}$ simulating nuclear density.} \label{plotcond}
\end{figure}

\vspace{0.3cm}In Fig. \ref{plotcond} we plot the lowest order
contribution to the DC conductivity as a function of temperature for
different choices of the scattering amplitudes that enter into the
pion width. We see that unitarization (resonances) makes the
conductivity grow from certain temperature. An increasing behavior
for the DC conductivity is also obtained in lattice calculations
\cite{Gupta}. The dots in the plot correspond to unitarizing the
scattering amplitudes at finite temperature, as we explained in
Section \ref{resonances}. We see that the thermal evolution of the
resonances does not affect much the conductivity. A more significant
effect on the conductivity is produced when finite nuclear density
is considered effectively by reducing $F_\pi$, but only at low
temperatures.

\vspace{0.3cm}As a phenomenological application of this result we can relate the electrical conductivity to the soft-photon spectrum emitted by the gas of pions produced after a Heavy-Ion Collision (HIC) \cite{FernandezFraile:05}. The rate of photons emerging from a thermal system is related to the EM current-current correlator by:
\begin{align}
\omega\frac{\mathrm{d}R_\gamma}{\mathrm{d}^3\bm{p}}=\frac{1}{8\pi^3}n_\mathrm{B}(\omega)\tensor{\rho}{^\mu_\mu}(\omega=|\bm{p}|)\ .
\end{align}
Now, using the Ward identity $p_\mu\rho^{\mu\nu}=0$, we can relate this rate to the conductivity as:
\begin{align}
\omega\frac{\mathrm{d}R_\gamma}{\mathrm{d}^3\bm{p}}(\omega\rightarrow 0^+,|\bm{0}|)=\frac{1}{4\pi^3}3T\sigma(T)\ .
\end{align}

\begin{figure}[h!]
\centerline{\resizebox{0.40\textwidth}{!}{\includegraphics{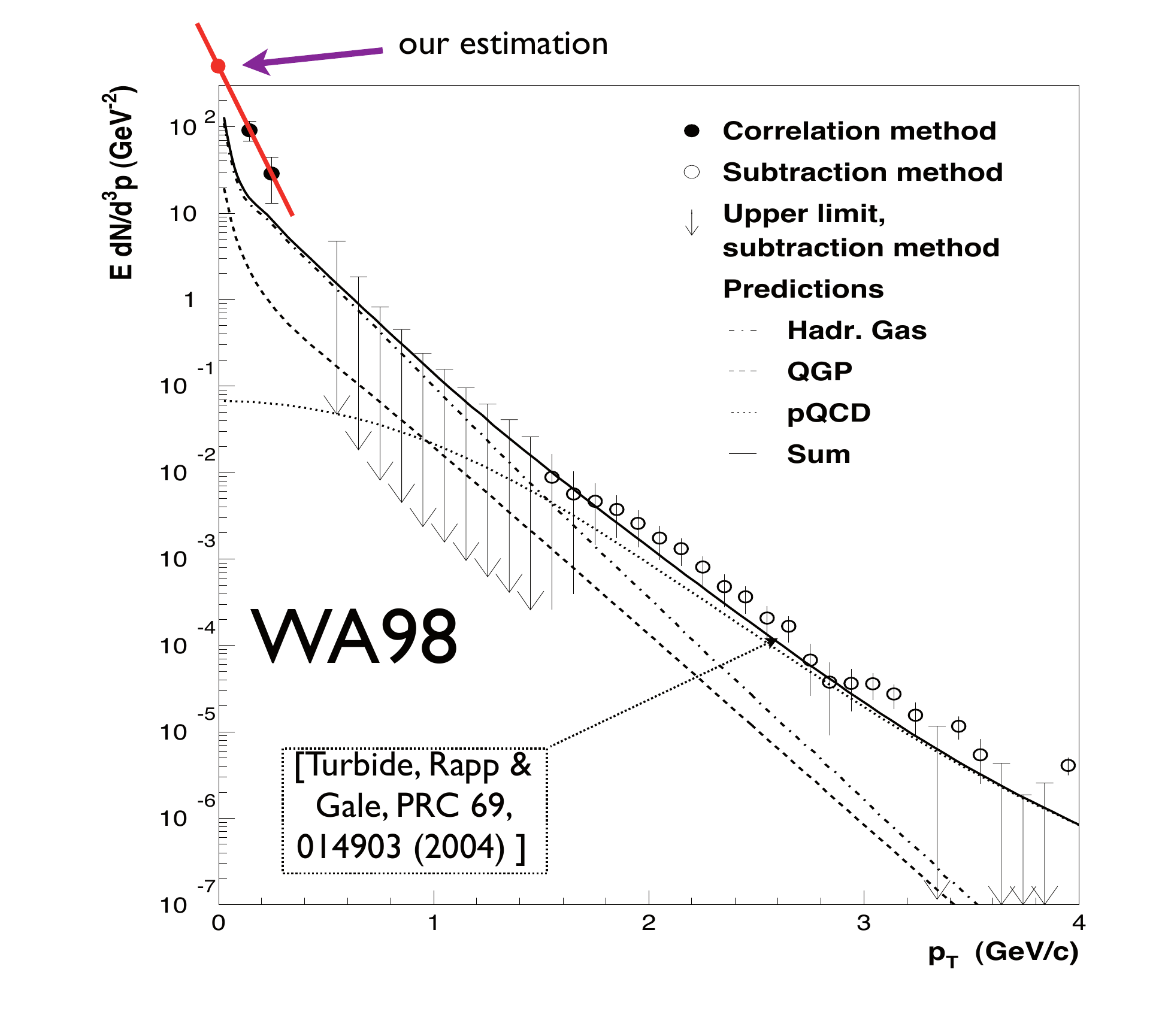}}}
\caption{Photon spectrum obtained by the experiment WA98 \cite{WA98}. We see that our estimate at the origin is compatible with a linear extrapolation from the two lowest-energy points.} \label{plotspectrum}
\end{figure}

So the DC conductivity is directly related to the soft-photon spectrum, i.e., photons emitted with almost zero momentum. In order to compare with experimental results we need to integrate this rate through the space-time evolution of the fireball produced after a HIC. For that purpose, we consider a simple hydrodynamical model of cylindrical symmetry in order to describe the expansion of the gas (Bjorken's model). Then, the measured rate would be given by:
\begin{align}
\omega\frac{\mathrm{d}N_\gamma}{\mathrm{d}^3\bm{p}}(p_T\rightarrow 0^+)\simeq\pi R_A^2\mathnormal{\Delta}\eta_\mathrm{nucl}\int\limits_{\tau_i}^{\tau_f}\frac{3T(\tau)\sigma(T(\tau))}{4\pi^3}\, \tau\mathrm{d}\tau
\ ,
\end{align}
where we consider lead-lead collisions at SPS energies in order to compare with the WA98 experiment \cite{WA98}, so $\sqrt{s}=158A\ \mathrm{GeV}$, the nuclei radius is $R_A\simeq 1.3\ \mathrm{fm}\ A^{1/3}\simeq 7.7\ \mathrm{fm}$, the expansion rapidity is $\mathnormal{\Delta}\eta_\mathrm{nucl}=2\acosh(\sqrt{s}/(2A\ \mathrm{GeV}))\\\simeq 10.1$, the initial proper time is $\tau_i\simeq 3\ \mathrm{fm}/c$, the final time $\tau_f\simeq 13\ \mathrm{fm}/c$, and we consider the cooling law of an ideal gas $T(\tau)=T_i(\tau_i/\tau)^{1/3}$ with $T_i\simeq 170\ \mathrm{MeV}$. With these parameters we obtain the estimate $\mathrm{d}N_\gamma/\mathrm{d}^3\bm{p}(p_T=0)\simeq 5.6\times 10^2$, that we indicate in the plot of the Fig. \ref{plotspectrum}. We see that other theoretical calculations do not fit the two lowest-energy points, while our result is compatible with a linear extrapolation to the origin from these two points. For low energies, the hadronic component is expected to dominate the spectrum, and in that regime a finite width in the particle propagator may be relevant due to the Landau-Pomeranchuk-Migdal effect, which was not taken into account in those studies and would make the contribution to the spectrum finite at the origin \cite{Rapp}.

\section{Thermal conductivity}

\begin{figure}[h!]
\centerline{\resizebox{0.40\textwidth}{!}{\includegraphics{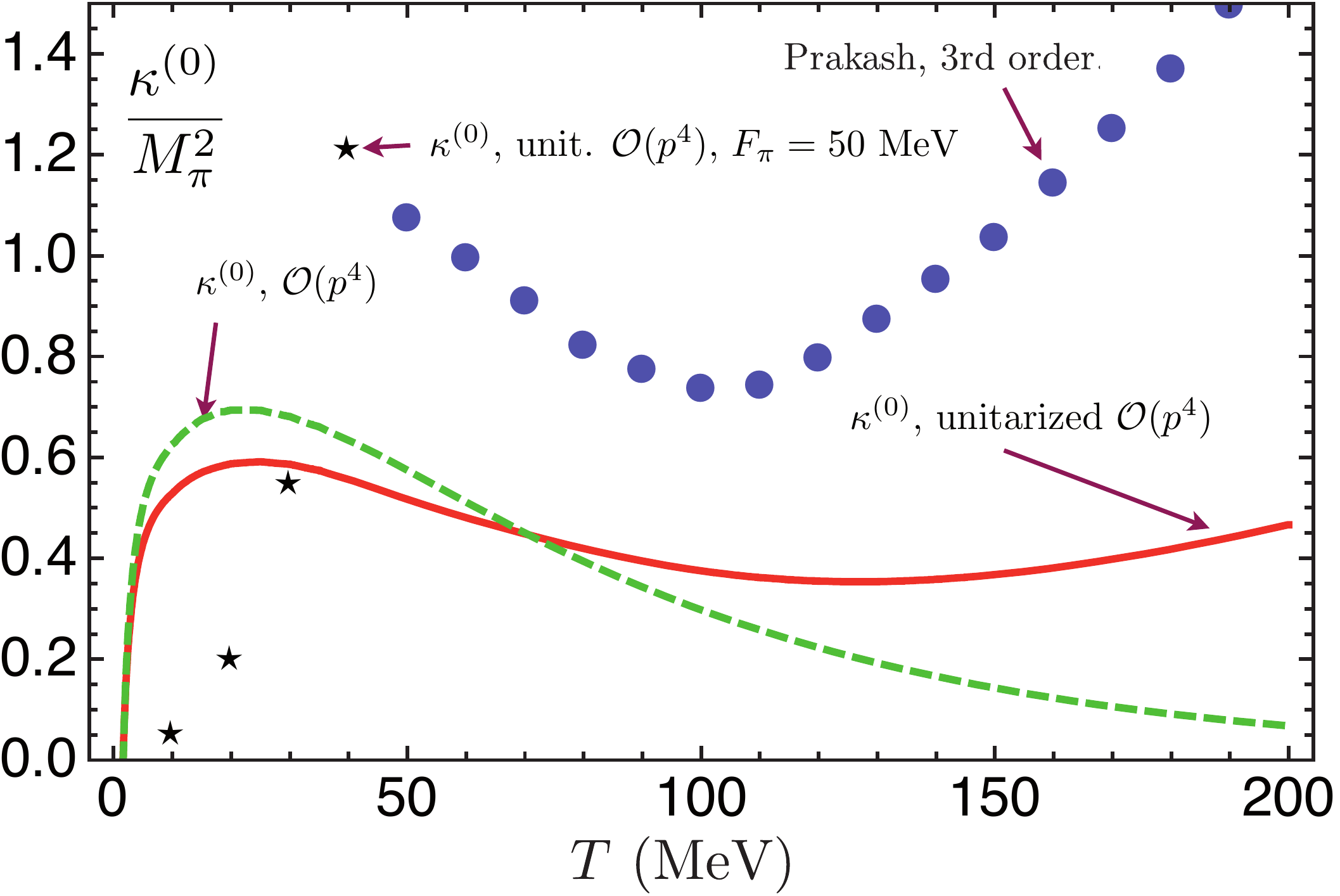}}}
\caption{Lowest-order contribution to the thermal conductivity as a function of the temperature. We compare with the analysis of \cite{Prakash}, which is based on kinetic theory.}
\label{plotthermal}
\end{figure}

Although in the pion gas the only strictly conserved quantity is
the energy-momentum tensor, in the energy and temperature regime
we are dealing with, it is a good approximation to assume
that $2\rightarrow 2$ collisions are the only relevant scattering
processes, which in practice means that the pion number is
approximately conserved \cite{Chemical}, yielding a nonzero
thermal conductivity even when $\mu=0$ \cite{Gavin:1985ph}.
Therefore, in order to compare with KT we apply
\begin{align}
\kappa=-\frac{\beta}{6}\lim\limits_{\omega\rightarrow 0^+}\lim\limits_{|\bm{p}|\rightarrow 0^+}\frac{\rho_\kappa(\omega,|\bm{p}|)}{\omega}\ ,
\end{align}
where
\begin{align}
\rho_\kappa(\omega,|\bm{p}|)=\int\mathrm{d}^4x\ \mathrm{e}^{\mathrm{i}p\cdot x}\langle[\hat{\mathcal{T}}^i(x),\hat{\mathcal{T}}_i(0)]\rangle\ ,
\end{align}
with $\mathcal{T}^{i}\equiv T^{i0}-hN^i$ \footnote{Note that in a previous calculation \cite{Florianopolis} we did not take into account the term with $N^i$, and therefore the result that we show here is qualitatively different at low temperatures.}. But now, under this assumption, thermal averages would only imply sums over states of well-defined number of particles, $|N\rangle$. Then, since in the diagram of Fig. \ref{leadingdiag} the energy-momentum enters to lowest order (i.e. without interaction), based on the KT theory expressions for the ideal-gas case \cite{deGroot}:
\begin{align}
T^{i0}&=\int\frac{\mathrm{d}^3\bm{p}}{(2\pi)^3}\ E_p v^i n_\mathrm{B}(E_p)\ ,\\
N^i&=\int\frac{\mathrm{d}^3\bm{p}}{(2\pi)^3}\ v^i n_\mathrm{B}(E_p)\ ,
\end{align}
with $v^i=p^i/E_p$, we define the operator $\hat{N}^i$ through its Feynman rule for the vertex in momentum space heuristically as $N^i\equiv T^{i0}/E_p$ (the limit of external momentum equal to zero is understood). According to this, the lowest-order contribution is then given by:
\begin{align}
\kappa^{(0)}=\frac{1}{8\pi^2T^2}\int\limits_0^\infty\mathrm{d}|\bm{p}|\ \frac{|\bm{p}|^4(E_p-h)^2}{E_p^2\mathnormal{\Gamma}_p} n_\mathrm{B}(E_p)[1+n_\mathrm{B}(E_p)]\ .
\end{align}
As we have seen in the derivation of the Kubo formula for the thermal conductivity, here $h$ represents the \emph{exact} heat function per particle. However, we will approximate in our results $h\equiv(\epsilon+P)/n=Ts/n$ ($s$ is the entropy density and $n$ the density of particles) by the corresponding ideal gas expression, which we expect to be reasonable for the temperatures considered here. For very low temperatures, $T\ll M_\pi$, we have:
\begin{align}\label{kappalowt}
\kappa^{(0)}\simeq 63\ \frac{T^{1/2}F_\pi^4}{M_\pi^{5/2}}\ .
\end{align}
By KT \cite{Landau}, $\kappa\sim T^{-1}(\bar{e}-h) l v$ ($\bar{e}$ is the mean energy per particle and $\displaystyle l\sim 1/(\sigma_{\pi\pi} n)$ is the particle mean free path). In the non-relativistic limit, $\bar{e}\sim M_\pi$, $h\sim 5T/2+M_\pi$, and then $\kappa\sim T^{1/2}$, so it is compatible with our result for low temperatures. In Fig. \ref{plotthermal} we compare our results for $\kappa$ with a KT analysis \cite{Prakash}. Again, unitarity changes the behavior of the transport coefficient with temperature. We now see that density effects modify significantly the thermal conductivity at high temperatures, as expected when introducing and additional conserved charge (the baryon number).

\section{Shear viscosity}
\begin{figure}[h!]
\centerline{\resizebox{0.40\textwidth}{!}{\includegraphics{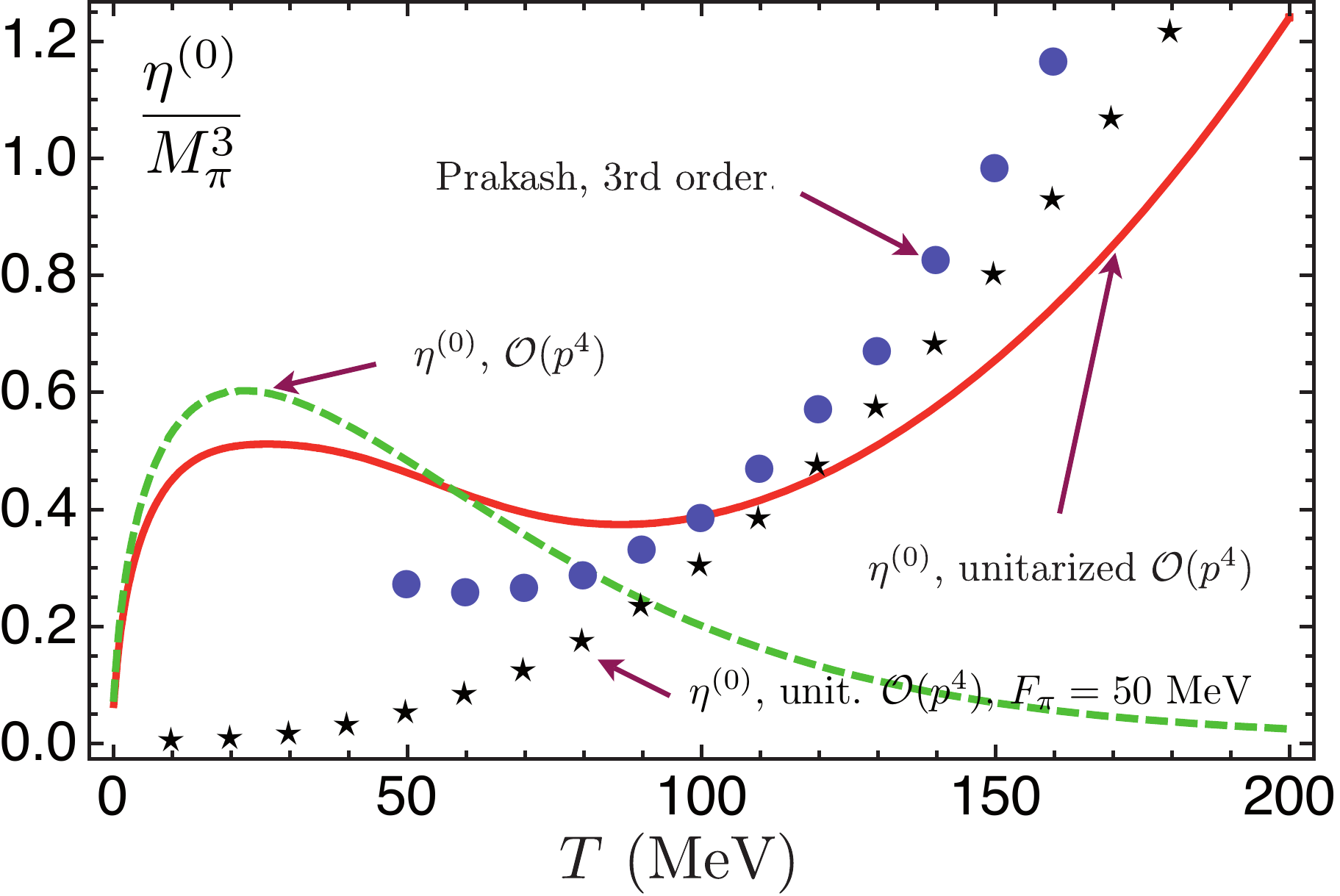}}}
\caption{Lowest-order contribution to the shear viscosity as a function of temperature for several approximations in the scattering amplitudes. We also compare with the KT results of \cite{Prakash} (dots).} \label{plotshear}
\end{figure}

It is given in LRT by:
\begin{align}
\eta=\frac{1}{20}\lim\limits_{\omega\rightarrow 0^+}\lim\limits_{|\bm{p}|\rightarrow 0^+}\frac{\rho_\eta(\omega,|\bm{p}|)}{\omega}\ ,
\end{align}
with
\begin{align}
\rho_\eta(\omega,|\bm{p}|)=\int\mathrm{d}^4x\ \mathrm{e}^{\mathrm{i}p\cdot x}\langle[\hat{\pi}^{ij}(x),\hat{\pi}_{ij}(0)]\rangle\ ,
\end{align}
and $\pi^{ij}\equiv T^{ij}-g^{ij}\tensor{T}{^k_k}/3$. Then, the lowest-order contribution is:
\begin{align}
\eta^{(0)}=\frac{1}{10\pi^2T}\int\limits_0^\infty\mathrm{d}|\bm{p}|\ \frac{|\bm{p}|^6}{E_p^2\mathnormal{\Gamma}_p} n_\mathrm{B}(E_p)[1+n_\mathrm{B}(E_p)]\ ,
\end{align}
For very low temperatures, $T\ll M_\pi$, we have:
\begin{align}\label{etalowt}
\eta^{(0)}\simeq 37\frac{T^{1/2}F_\pi^4}{M_\pi^{3/2}}\ .
\end{align}
In Fig. \ref{plotshear} we compare our results for viscosities with those obtained by Prakash \textit{et al.} using a KT analysis \cite{Prakash}. We also agree with a work by Dobado \textit{et al.} \cite{DobadoFelipe} for the pion gas in KT. We see that nuclear density effects would only imply a significant change in the shear viscosity at low temperatures. By non-relativistic KT we expect the behavior $\eta,\zeta\sim M_\pi v n l$, thus $\eta,\zeta\sim\sqrt{T}$ and both viscosities should then be of the same order at very low temperatures.

\begin{figure}[h!]
\centerline{\resizebox{0.40\textwidth}{!}{\includegraphics{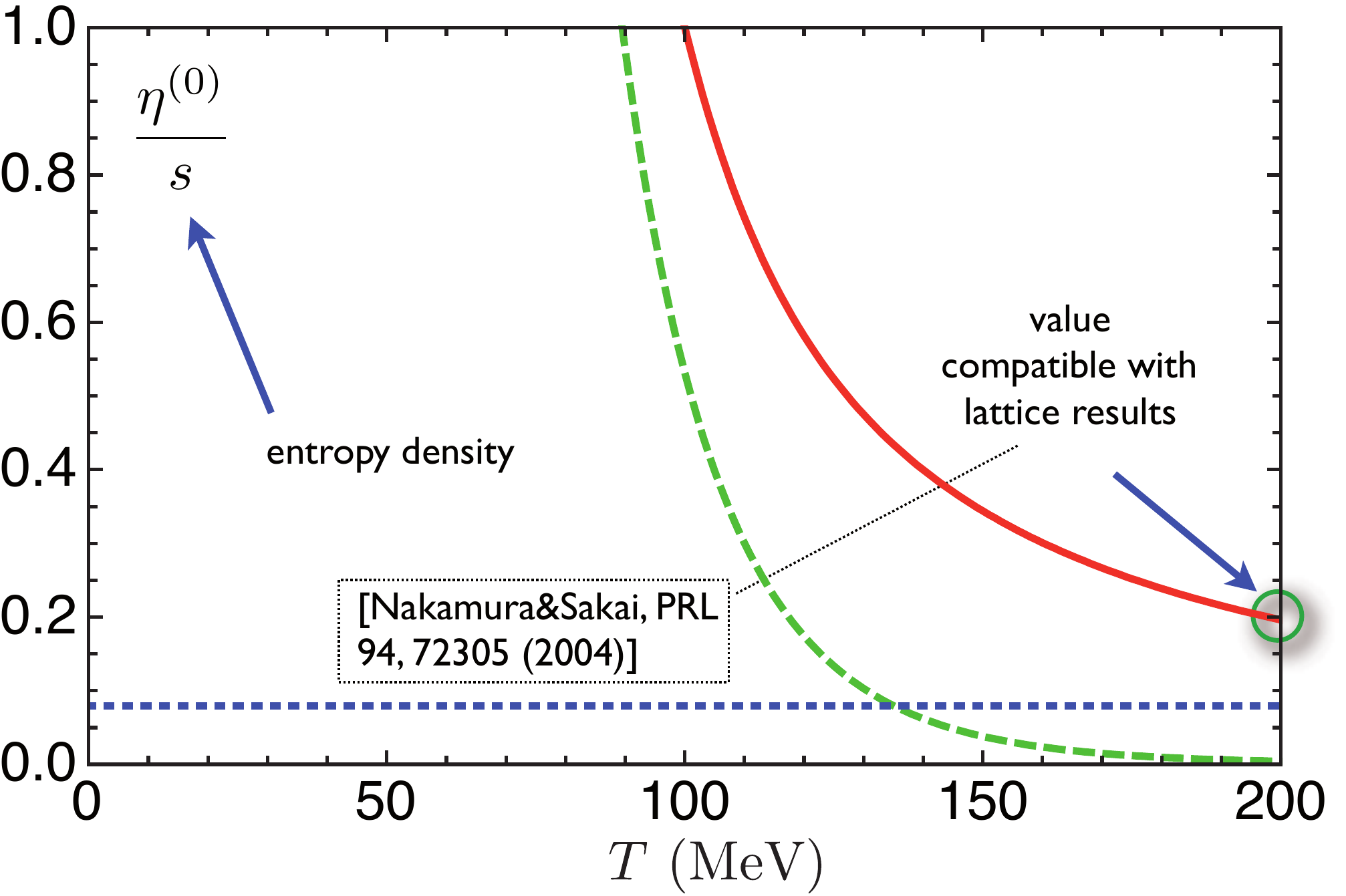}}}
\caption{Shear viscosity over the entropy density as a function of temperature. The horizontal dashed line corresponds to the AdS/CFT bound $1/(4\pi)$.} \label{plotetas}
\end{figure}

\vspace{0.3cm}Unitarity makes the quotient $\eta/s$ ($s$ is the entropy density) for the pion gas respect the bound $1/(4\pi)$ predicted by Kovtun \textit{et al.} \cite{Kovtun}, as we can see in Fig. \ref{plotetas}. Without unitarization the Uncertainty Principle would be also violated eventually, since $\eta/s\sim\epsilon\tau/n\sim E\tau\gtrsim 1$. Furthermore, near $T_c$ our value for
$\eta/s$ is not far from recent lattice and model estimates \cite{Nakamura}. Although we do not represent it in the figure, we do obtain a behavior for $\eta/s$ growing slowly with $T$ for temperatures (unrealistic) $>550\ \mathrm{MeV}$. A slowly increasing behavior is also obtained by calculations from the quark gluon plasma (QGP) phase \cite{Kapusta}, although a recent work predicts a more pronounced increase near the phase transition, in the so-called semi-QGP phase \cite{Yoshimasa}. As another check, we can compute the sound attenuation length, which is given by (neglecting the contribution from the bulk viscosity) $\mathnormal{\Gamma}_\mathrm{s}\simeq 4\eta/(3sT)$, and is directly related to phenomenological effects such as \emph{elliptic flow} or \emph{HBT radii}. We get, at $T=180\ \mathrm{MeV}$, the value $\mathnormal{\Gamma}_\mathrm{s}\simeq 0.55\ \mathrm{fm}$, in agreement with the estimate by Teaney \cite{Teaney}.

In the chiral limit with only $\mathcal{O}(p^2)$, we get exactly
$\eta=18\pi\zeta(3) F_\pi^4/(25 T)$ with the Riemann's zeta
function $\zeta(3)\simeq 1.2$. This $1/T$ decreasing behaviour,
obtained for instance in \cite{chennakano07}, would imply that the
AdS/CFT bound for $\eta/s$ is violated at some point, departing
also from the phenomenological estimates discussed above. This
highlights the importance of reproducing correctly the high energy
features of particle scattering, as we do within our unitarized
approach.
\section{Bulk viscosity}
As we have seen, the expression for the bulk viscosity in LRT is:
\begin{align}
\zeta=\frac{1}{2}\lim\limits_{p^0\rightarrow 0^+}\lim\limits_{|\bm{p}|\rightarrow 0^+}\frac{\rho_\zeta(p^0,|\bm{p}|)}{p^0}\ ,
\end{align}
with
\begin{align}
\rho_\zeta(p^0,|\bm{p}|)=\int\mathrm{d}^4x\ \mathrm{e}^{\mathrm{i}p\cdot x}\langle[\hat{\mathcal{P}}(x),\hat{\mathcal{P}}(0)]\rangle\ ,
\end{align}
where, if there is no conserved charge in the system, $\mathcal{P}\equiv-\tensor{T}{^k_k}/3-c_s^2 T_{00}$. Then, the lowest order contribution to the bulk viscosity, $\zeta^{(0)}$ corresponding to the diagram of Figure \ref{leadingdiag} is:
\begin{align}\label{zetalowest}
\zeta^{(0)}=&\ \frac{3}{4\pi^2 T}\int\limits_0^\infty\mathrm{d}|\bm{p}|\ \frac{|\bm{p}|^2(|\bm{p}|^2/3-c_s^2E_p^2)^2}{E_p^2\mathnormal{\Gamma}_p}\notag\\
&\times n_\mathrm{B}(E_p)[1+n_\mathrm{B}(E_p)]\ ,
\end{align}
For very low temperatures, $T\ll M_\pi$, the leading-order contribution simplifies:
\begin{align}\label{zetalowt}
\zeta^{(0)}\simeq 13\frac{T^{1/2}F_\pi^4}{M_\pi^{3/2}}\ .
\end{align}

In the chiral limit, $M_\pi=0$, we have the simple relation between the shear and bulk viscosities, $\zeta^{(0)}=15(1/3-c_s^2)^2\eta^{(0)}$, in agreement (parametrically) with the result obtained in the high-temperature regime of QCD \cite{Dogan}. This result implies that the bulk viscosity is suppressed with respect to the shear viscosity at large temperatures, as a consequence of conformal invariance, since $c_s^2=1/3$ for a free massless gas and, in fact,  the bulk viscosity vanishes exactly for a conformally invariant theory. However, it also suggests that  conformal breaking, as in the case of QCD through explicit and anomalous terms (see below) may induce sizable values for $\zeta/\eta$, which would have interesting phenomenological consequences, since bulk viscosity is generically assumed to be negligible. This observation, supported by recent QCD analyses (see below) has led us to analyze recently in \cite{ffgn08} the correlation between conformal invariance and the bulk viscosity in the pion gas within the unitarized ChPT context. We reproduce some of the main results of that paper here, with more detail and emphasizing some quantitative aspects of the analysis.

\vspace{0.3cm}In the lowest-order contribution (\ref{zetalowest}), $c_s$ is the exact (non-perturbative) speed of sound of the pion gas. However, we can only calculate it within some approximation. In order to estimate the speed of sound we will analyze the relation between the trace anomaly and the bulk viscosity \cite{Kharzeev,Karsch}. The scale invariance of the QCD lagrangian is broken explicitly by the quark mass and  by the running of the strong coupling constant at the quantum level \cite{Collins}:
\begin{align}
\partial_\mu s^\mu=\tensor{T}{^\mu_\mu}=\frac{\beta(g)}{2g}F_{\mu\nu}^a F_a^{\mu\nu}+\{1+\gamma(g)\}m\bar{q}q\ ,
\end{align}
where $s^\mu=T^{\mu\nu}x_\nu$ is the dilation current, $\beta(g)$ is the $\beta$-function, $\gamma(g)$ is the anomalous dimension of the quark mass, and we consider the case of two flavors with $m\equiv m_\mathrm{u}=m_\mathrm{d}$. At finite temperature, the average of the trace anomaly is given in terms of thermodynamical quantities, $\langle\theta\rangle_T\equiv\langle\tensor{T}{^\mu_\mu}\rangle_T=\epsilon-3P$, and it has already been calculated on the lattice for the pure glue theory \cite{Boyd} as well as for QCD with almost physical quark masses \cite{Cheng}. In Refs. \cite{Kharzeev} and \cite{Karsch} it was found through a low-energy theorem of QCD, and assuming some reasonable ansatz for the spectral function, a relation between the trace anomaly and the bulk viscosity. For instance, in the chiral limit, this relation reads \cite{Kharzeev}:
\begin{align}
\int\limits_{-\infty}^{\infty}\mathrm{d}\omega\ \frac{\rho_{\theta\theta}(\omega,0)}{\omega}&=-\left(4-T\frac{\partial}{\partial T}\right)\langle\theta\rangle_T\notag\\
&=T^5\frac{\partial}{\partial T}\frac{(\epsilon-3P)^*}{T^4}+16|\epsilon_v|\ ,
\end{align}
where $(\cdot)^*\equiv(\cdot)_T-(\cdot)_0$ is what is measured on the lattice, and $|\epsilon_v|$ is the energy density of the vacuum. The spectral density $\rho_{\theta\theta}$ involves the correlator of the trace of the energy-momentum tensor, and it is related to the pressure-pressure correlator by \cite{Meyer,Pica}:
\begin{align}
\rho_{\theta\theta}(\omega,0)=9\rho_{PP}(\omega,0)+\omega\delta(\omega)T\partial_T(\epsilon-6P)\ .
\end{align}

\begin{figure}[h!]
\centerline{\resizebox{0.35\textwidth}{!}{\includegraphics{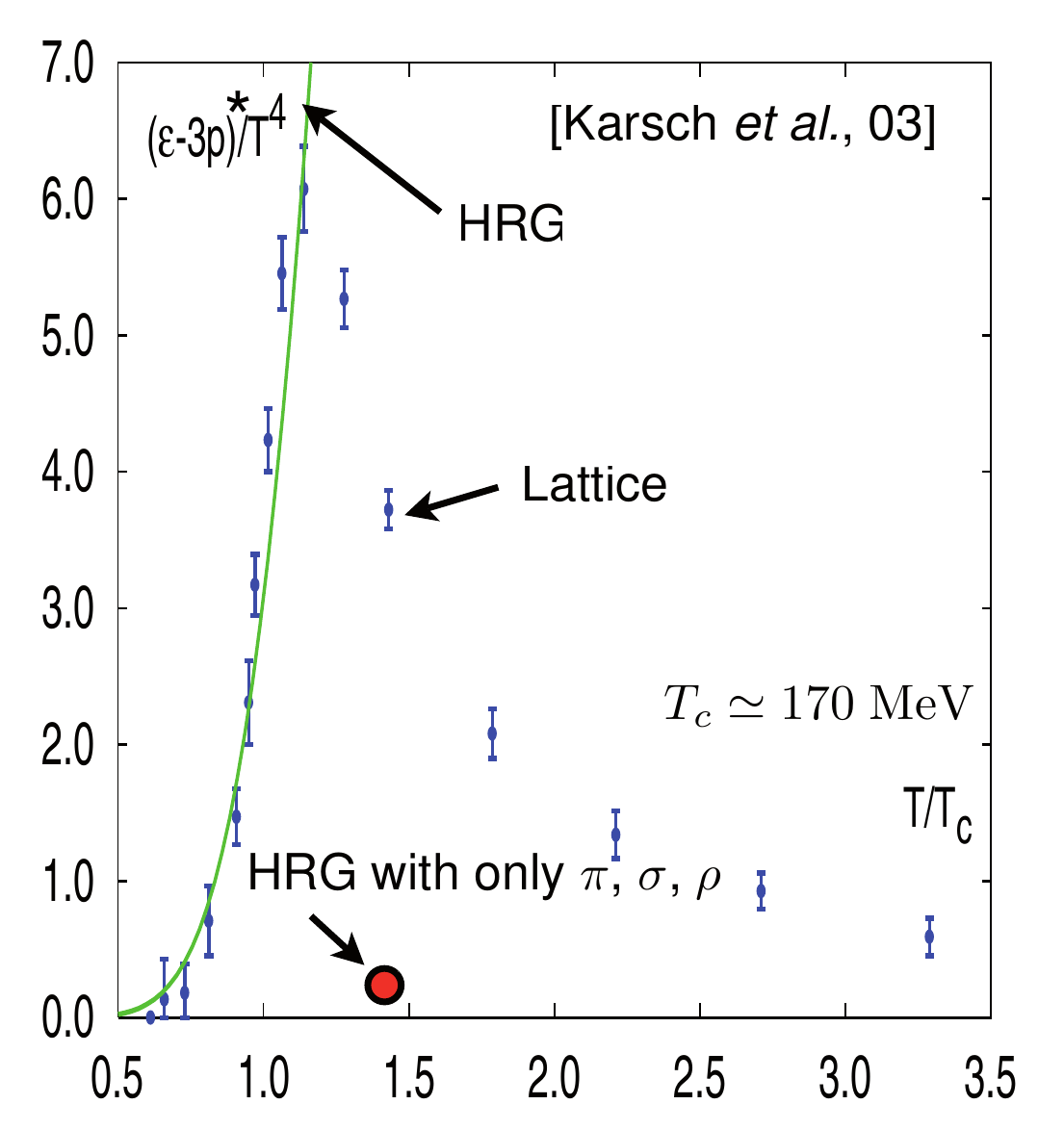}}}
\caption{Interaction measure calculated on the lattice (dots) and in the HRG approximation (green line, taking into account 1026 states in total, see text), from \cite{Karsch:03}. The big red dot corresponds to the result from the HRG approximation taking into account only pions, the $f_0(600)$ and the $\rho(770)$ states.} \label{traceanomHRG}
\end{figure}

\begin{figure}[h!]
\centerline{\resizebox{0.40\textwidth}{!}{\includegraphics{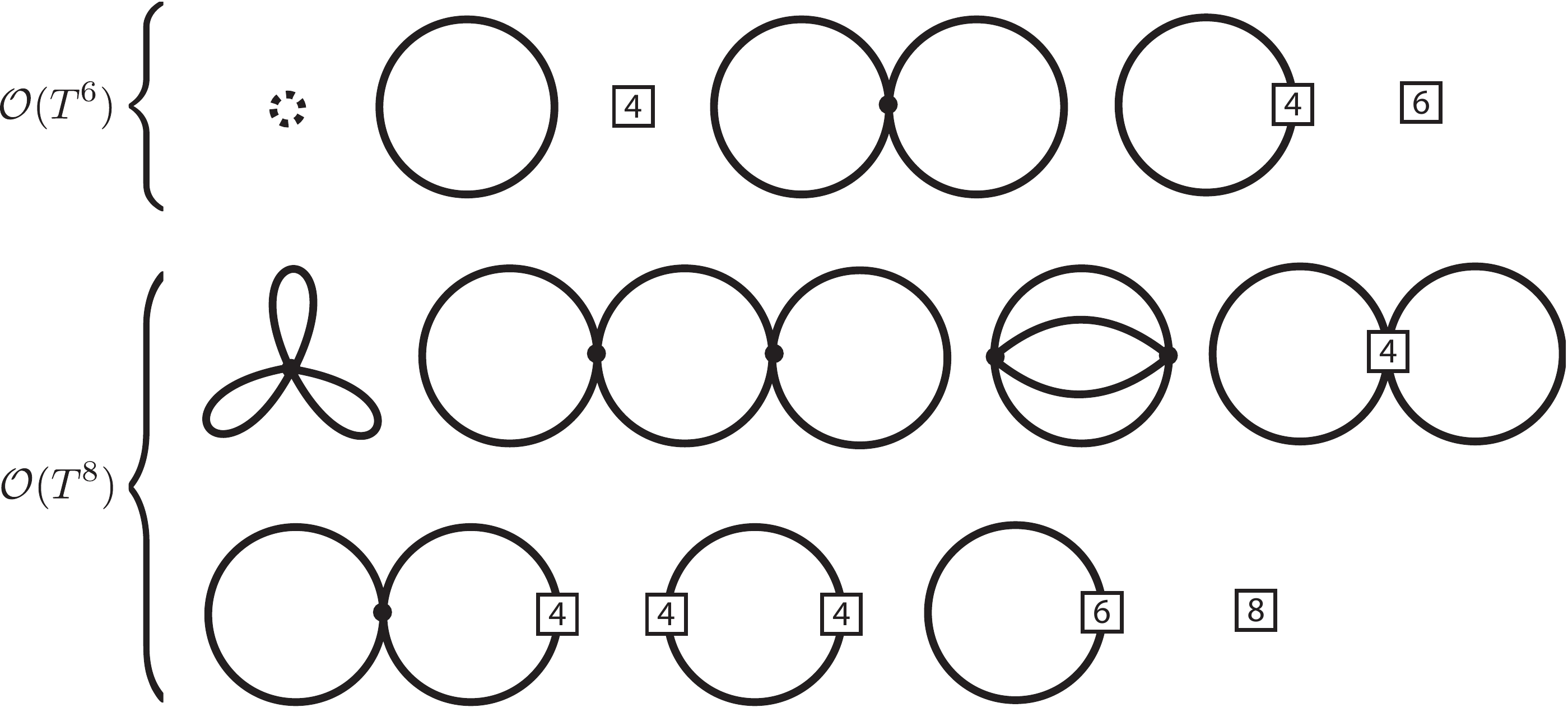}}}
\caption{Diagrams contributing to order $\mathcal{O}(T^6)$ and $\mathcal{O}(T^8)$ to the partition function of the pion gas. The first diagram just corresponds to a constant term in the lagrangian, boxes labeled with the number $n$ correspond to counter-terms from the lagrangian $\mathcal{L}_n$.} \label{diagsGL}
\end{figure}

Now, since the interaction measure, defined as $\mathnormal{\Delta}\equiv\langle\theta\rangle^*/T^4$, has a peak near the critical temperature (see Fig. \ref{traceanomHRG}), it is reasonable to assume the following ansatz for the spectral density of the pressure-pressure correlator \cite{Kharzeev,Karsch,Pica}:
\begin{align}
\rho_{PP}(\omega,0)=\frac{\zeta}{\pi}\frac{\omega\omega_0^2}{\omega_0^2+\omega^2}\ ,
\end{align}
where $\omega_0\sim 1\ \mathrm{GeV}$. So a maximum in the interaction measure would imply a maximum in the bulk viscosity. Motivated by this, in order to obtain a good estimation of the speed of sound, we first calculate the interaction measure in ChPT. The interaction measure can be calculated from the pressure through
\begin{align}
\langle\tensor{T}{^\mu_\mu}\rangle_T=T^5\frac{\mathrm{d}}{\mathrm{d}T}\left(\frac{P}{T^4}\right)\ .
\end{align}
A calculation in ChPT of the pion gas pressure up to 3-loop order by Gerber and Leutwyler \cite{Gerber} is available. The diagrams contributing to the partition function of the pion gas are represented in Fig. \ref{diagsGL} up to order $\mathcal{O}(T^8)$ according to the chiral counting.

\begin{figure}[h!]
\centerline{\resizebox{0.40\textwidth}{!}{\includegraphics{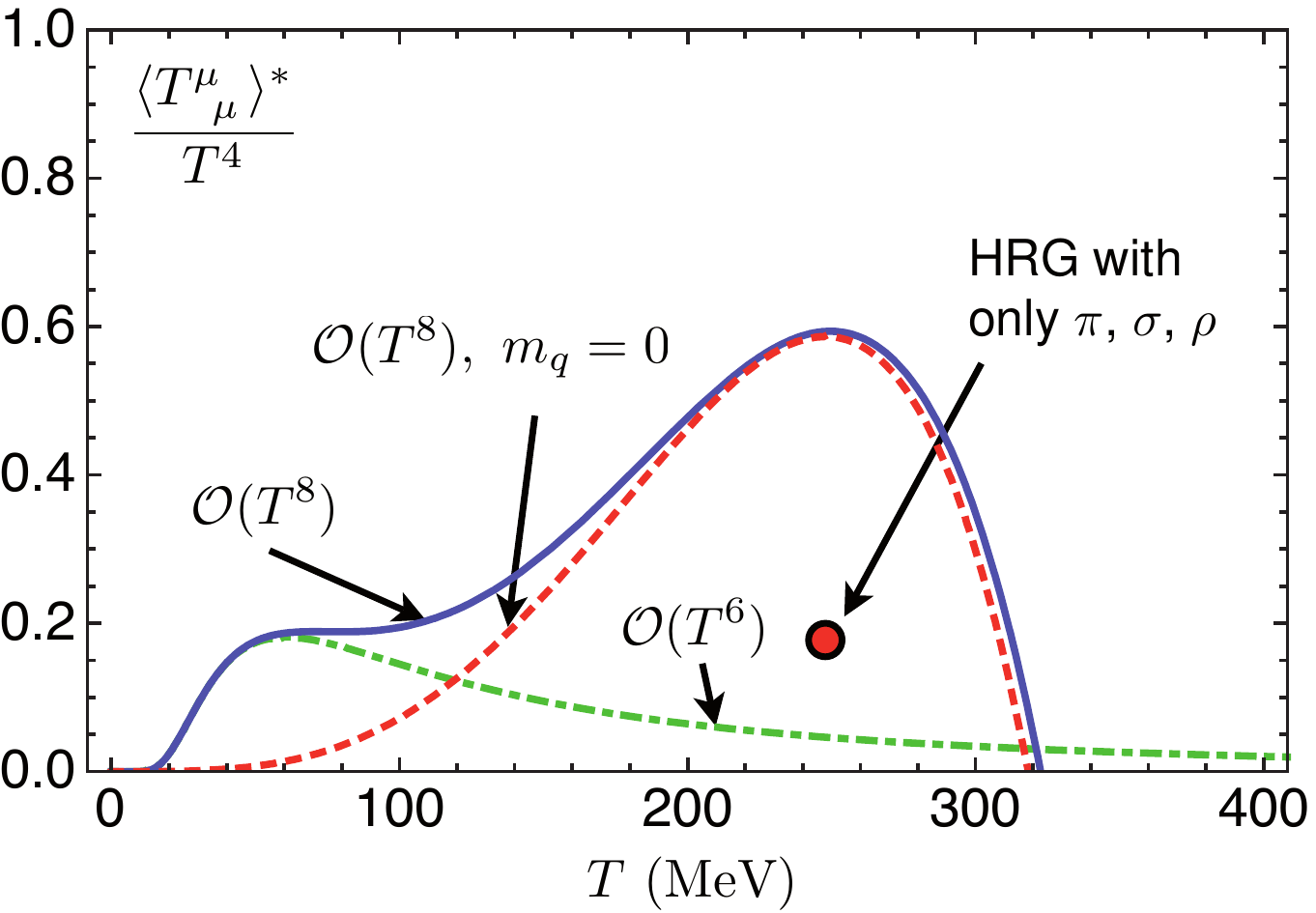}}}
\caption{Invariant measure calculated perturbatively in ChPT for the pion gas. Each order of approximation corresponds to a definite set of diagrams in the partition function, see Fig. \ref{diagsGL}. The big red dot corresponds to the HRG approximation taking into account only the pion, the $f_0(600)$ and $\rho(770)$ states (see text).} \label{traceanomChPT}
\end{figure}

Using the result for the pressure from \cite{Gerber}, in Fig.
\ref{traceanomChPT} we plot the interaction measure for several
orders in the pressure. The first peak corresponds to the explicit
breaking of scale symmetry by the quark mass and it has also been
obtained in other works \cite{Li:2008zp}. Interestingly, at order
$\mathcal{O}(T^8)$ there appears another peak near the phase
transition, that would correspond to the gluon condensate
contribution. We then see that, in the chiral limit, interaction
kicks in at $\mathcal{O}(T^8)$ (the $\mathcal{O}(T^6)$ diagrams
would only imply a renormalization of the mass in this case). For
$M_\pi=0$, the trace anomaly has a simple expression at this order
\cite{Gerber,Leutwyler}:
\begin{align}
\langle\tensor{T}{^\mu_\mu}\rangle^*=\frac{\pi^2}{270}\frac{T^8}{F_\pi^4}\left(\ln\frac{\mathnormal{\Lambda}_p}{T}-\frac{1}{4}\right)\ ,
\end{align}
with $\mathnormal{\Lambda}_p\sim400\ \mathrm{MeV}$ for our choice of $\bar l_i$ given in Section \ref{resonances}. The result is very dependent on the choice of $\bar l_i$, which in our case encode important non-perturbative information, as discussed in \cite{ffgn08}. However, we see that the result of this perturbative calculation for the pion gas still is a factor 10 smaller than the lattice and full-HRG results, see Fig.
\ref{traceanomHRG}.

\begin{figure}[h!]
\centerline{\resizebox{0.35\textwidth}{!}{\includegraphics{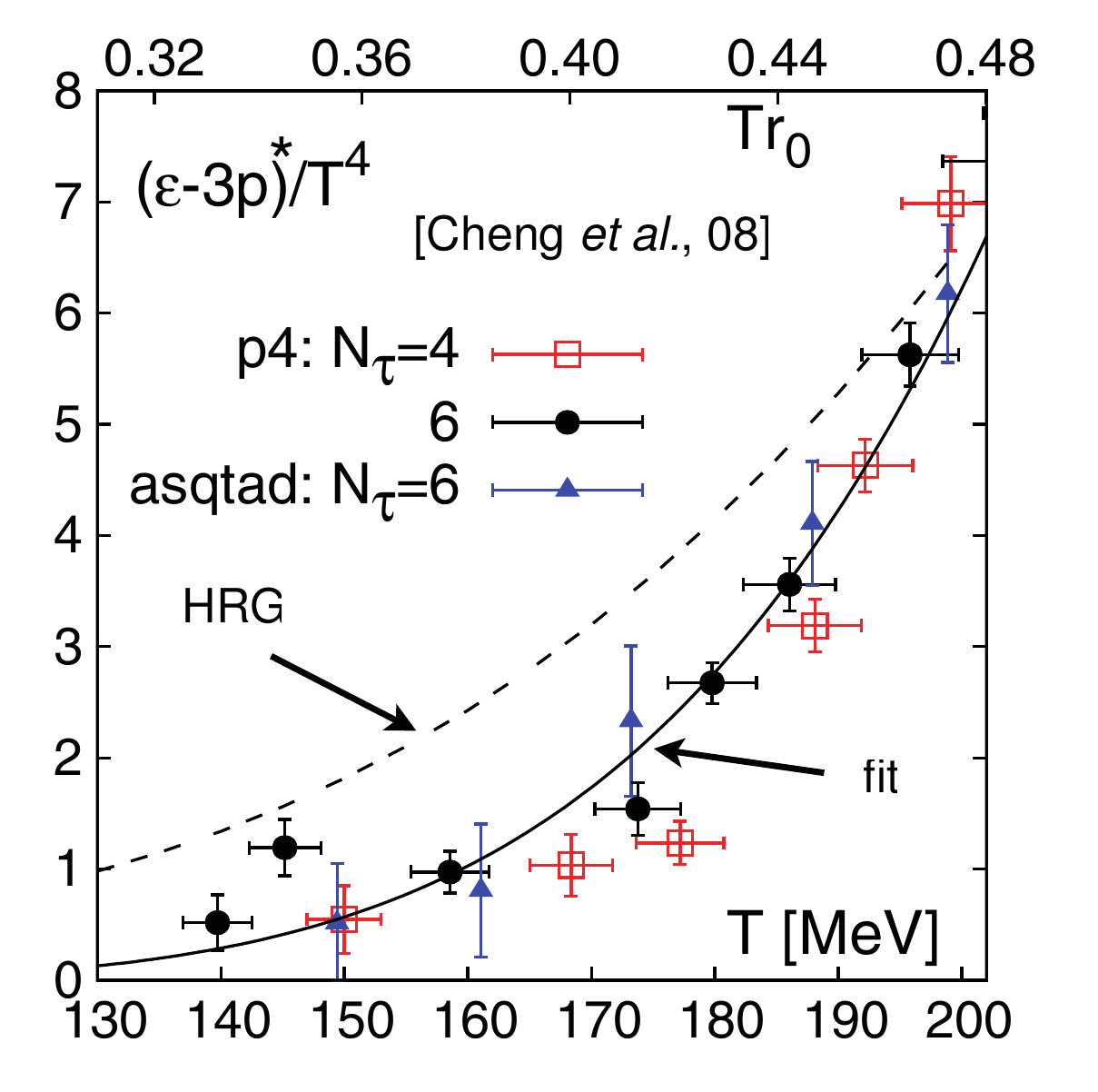}}}
\caption{Comparison between the lattice results (dots and continuous line) with almost physical quark masses and the HRG approximation (dashed line) for temperatures below the phase transition, from \cite{Cheng}.} \label{traceanomlattice}
\end{figure}

The Hadron Resonance Gas (HRG) approximation considers a free (non-interacting) gas which consists of all the baryonic and mesonic states up to $2\ \mathrm{GeV}$, 1026 in total \cite{Karsch:03}. Therefore, in this approximation the interaction measure is given by:

\begin{align}
\mathnormal{\Delta}&\equiv\frac{(\epsilon-3P)^*}{T^4}=\sum\limits_{i=1}^{1026}\frac{(\epsilon_i-3P_i)^*}{T^4}\notag\\
&=\sum\limits_{i=1}^{1026}\frac{g_i}{2\pi^2}\sum\limits_{k=1}^\infty\eta^{k+1}\frac{(\beta m_i)^3}{k}K_1(k\beta m_i)\ ,
\end{align}
where $g_i$ denotes the degeneracy of the state, $\eta=\pm 1$ depending on whether it is a boson or a fermion respectively, and $K_1$ is the modified Bessel function of the second kind. In Fig. \ref{traceanomlattice}, from \cite{Cheng}, the lattice results for the interaction measure are compared with the HRG results for temperatures below the phase transition. We see that the HRG approximation fits better the lattice results near the maximum of the peak, while as the temperature decreases they start to separate from each other. This might be due to the values of the quark masses taken in \cite{Cheng}, since the HRG approximation and ChPT should coincide for very low temperatures. It is important to remark that although the HRG approximation gives a value for the interaction measure compatible with lattice results near the peak, it is a monotonously increasing curve, so it does not have the shape of a peak, which the ChPT calculation does have because it includes the interaction between the Goldstone bosons (in the HRG the non-zero interaction measure comes from the explicit breaking due to the non-zero mass of the states).

We are also interested in  the influence of the in-medium $f_0(600)$ and $\rho(770)$ resonances on the trace anomaly and eventually on the bulk viscosity. In order to do it, we calculate the interaction measure in the Virial Gas Approximation (VGA) which allows us to introduce the unitarized scattering amplitudes in the dilute gas regime. According to the VGA, to the lowest order in the interaction, the pressure of the gas is given by
\cite{Ruben}:

\begin{align}
\beta P=\sum\limits_i\left(B_i^{(1)}\xi_i+B_i^{(2)}\xi_i^2+\sum_{j\geq i}B_{\text{int}}\xi_i\xi_j+\ldots\right)\ ,
\end{align}
where
\begin{align}
B_i^{(n)}=&\ \frac{g_i\eta_i^{n+1}}{2\pi^2n}\int\limits_0^\infty\mathrm{d}p \ p^2\mathrm{e}^{-n\beta(E_i-m_i)}\ ,
\end{align}

\begin{figure}[h!]
\centerline{\resizebox{0.40\textwidth}{!}{\includegraphics{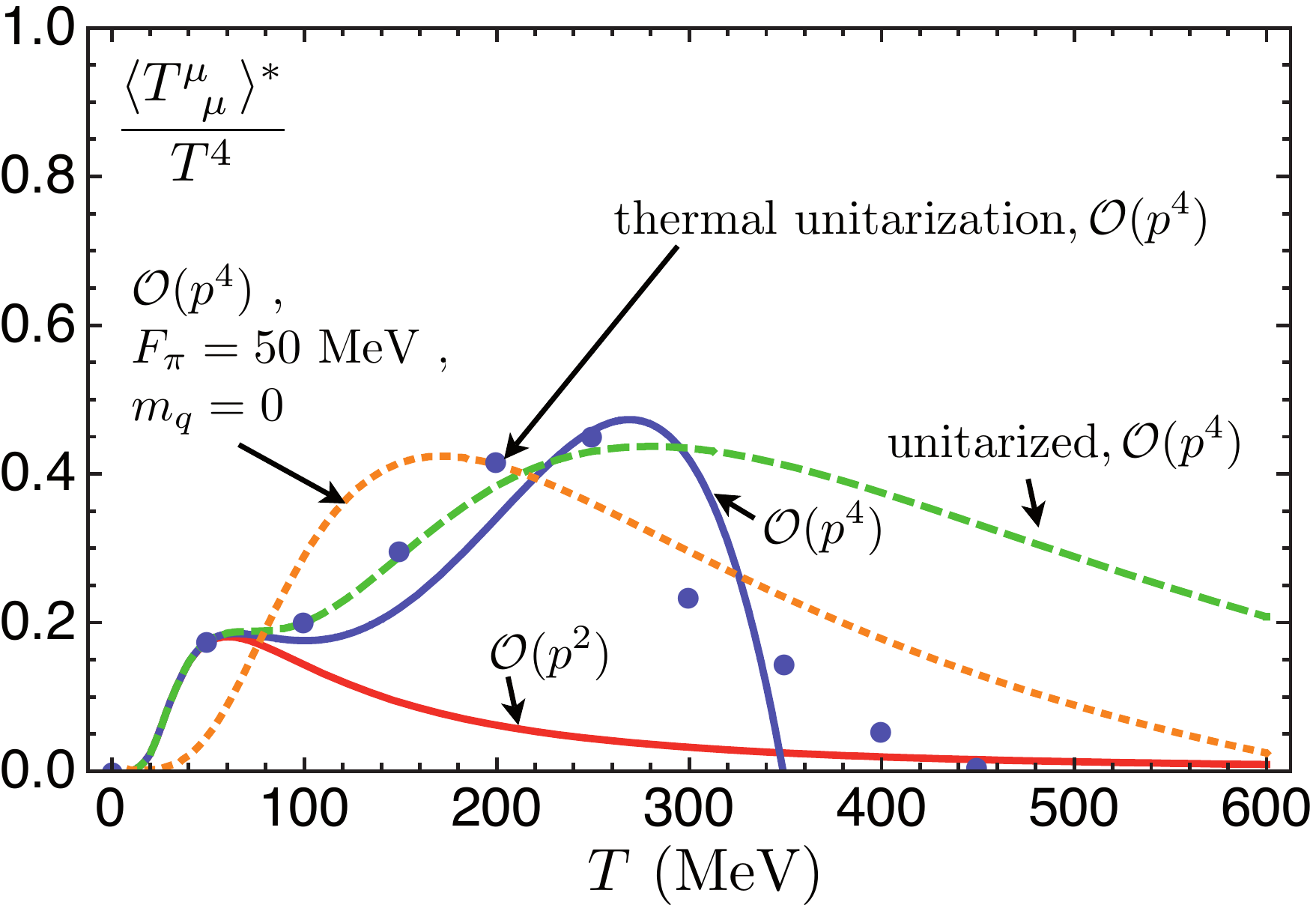}}}
\caption{Invariant measure in the VGA approximation (see text) for several scattering amplitudes. The in-medium evolution of the $f_0(600)$ and $\rho(770)$ resonances do no change significantly the height of the interaction peak.} \label{traceanomVG}
\end{figure}

\vspace{-0.5cm}\begin{align}
B_{ij}^{\mathrm{int}}=&\ \frac{\mathrm{e}^{\beta(m_i+m_j)}}{2\pi^3}\int\limits_{m_i+m_j}^\infty\mathrm{d}E\ E^2K_1(\beta E)\\\notag
&\times\sum\limits_{I,J,S}(2I+1)(2J+1)\delta_{IJS}^{ij}(E)\ ,
\end{align}
with $\xi_i\equiv\mathrm{e}^{\beta(\mu_i-m_i)}$, and $\delta_{IJS}^{ij}$ are the phase-shifts. In Fig. \ref{traceanomVG} we plot the interaction measure for several scattering amplitudes in the VGA. The interaction peak is obtained when considering $\mathcal{O}(p^4)$ amplitudes, and its height is almost equal to the one obtained with the perturbative calculation of Fig. \ref{traceanomChPT}. We then observe that the in-medium evolution of the $f_0(600)$ and $\rho(770)$ resonances (see Section \ref{resonances}) does not change significantly the height of the interaction peak.

According to this, we would not expect a big change in the bulk viscosity due to the in-medium evolution of resonances. In fact, this is consistent with the idea \cite{ffgn08} that chiral restoration is not the main source for the conformal anomaly peak described above (unlike the vanishing temperature of the chiral condensate in ChPT, the position of this peak does not change in the chiral limit). Nevertheless, a recent work \cite{Paech} based on the Linear Sigma Model, obtains a peak in the bulk viscosity at the chiral phase transition due to a minimum in the sigma mass, since in this model $\zeta\propto\mathnormal{\Gamma}_\sigma/m_\sigma^2$.

\begin{figure}[h!]
\centerline{\resizebox{0.23\textwidth}{!}{\includegraphics{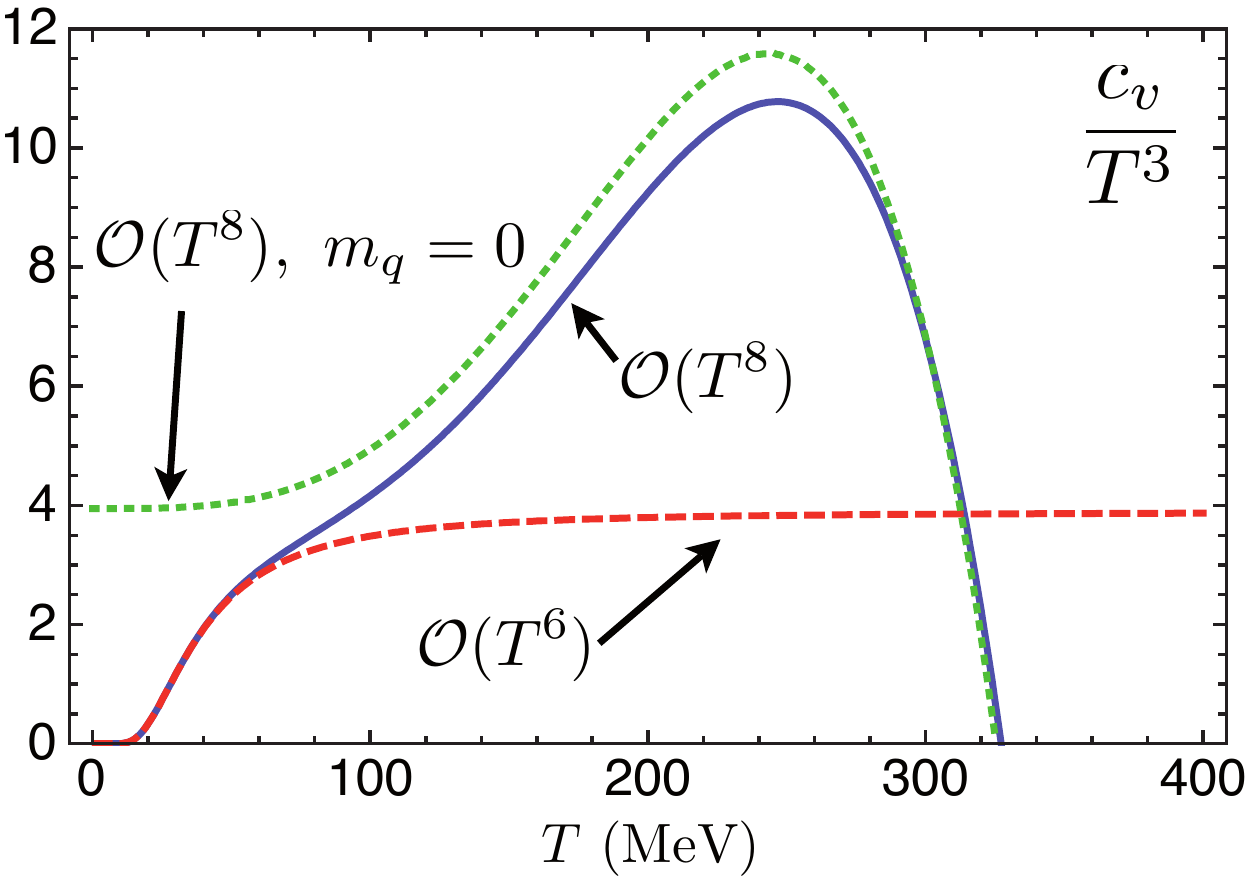}}\hspace{0.3cm}\resizebox{0.23\textwidth}{!}{\includegraphics{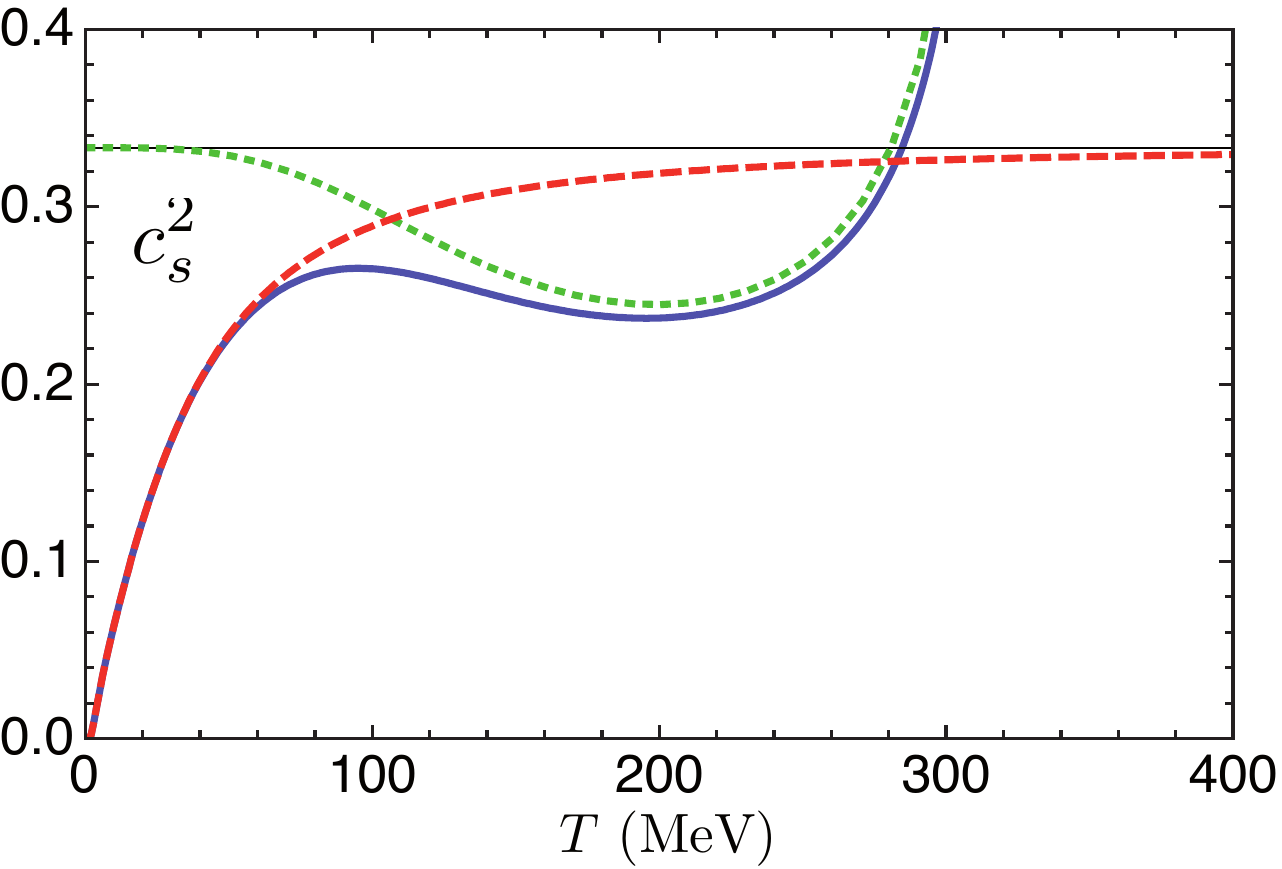}}}
\caption{Heat capacity (left) and speed of sound squared (right) for the pion gas in ChPT from the pressure obtained in \cite{Gerber}. At $\mathcal{O}(T^8)$, a maximum (minimum) in the heat capacity (speed of sound squared) is obtained near the chiral phase transition temperature.} \label{plotcvcs2}
\end{figure}
\begin{figure}[h!]
\centerline{\resizebox{0.35\textwidth}{!}{\includegraphics{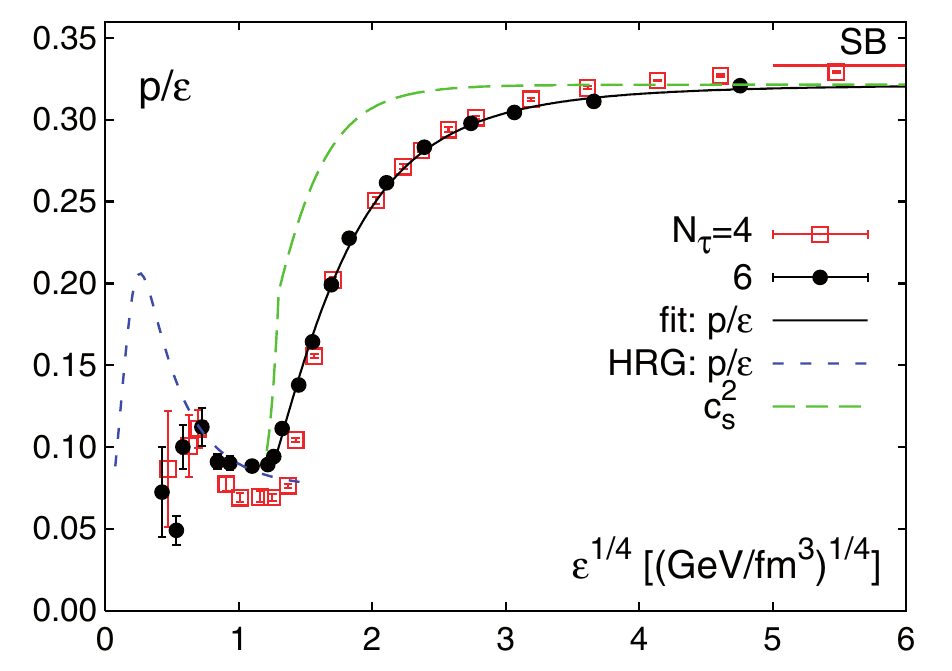}}}
\caption{Lattice result with almost physical quark masses for the speed of sound squared and the equation of state, from \cite{Cheng} ($N_f=3$).} \label{plotcs2lattice}
\end{figure}

\vspace{0.3cm}From the pressure we can also calculate other
thermodynamical quantities like the entropy density, $s=\partial
P/\partial T$, the specific heat, $c_v=\partial\epsilon/\partial
T=T\partial s/\partial T$, and the speed of sound, $c_s^2=\partial
P/\partial\epsilon=s/c_v$. In Fig. \ref{plotcvcs2} we plot the
specific heat and speed of sound squared for the pion gas in ChPT
for several approximations. We see that the maximum at the
interaction measure implies a minimum in the speed of sound, so
looking at the expression (\ref{zetalowest}) we then expect a
maximum in the bulk viscosity at the corresponding temperature. In
Fig. \ref{plotcs2lattice} we also show the speed of sound squared
and the equation of state as a function of the energy density
obtained in the lattice with almost physical quark masses
\cite{Cheng}. We see that the minimum in the speed of sound
squared for the pion gas is still a factor 2.5 less deep than the
value from the lattice, but we have to bear in mind that we are
dealing with a $m\neq 0$ two-flavor approximation, whose critical behavior
should be that of a $O(4)$-crossover.

\begin{figure}[h!]
\centerline{\resizebox{0.40\textwidth}{!}{\includegraphics{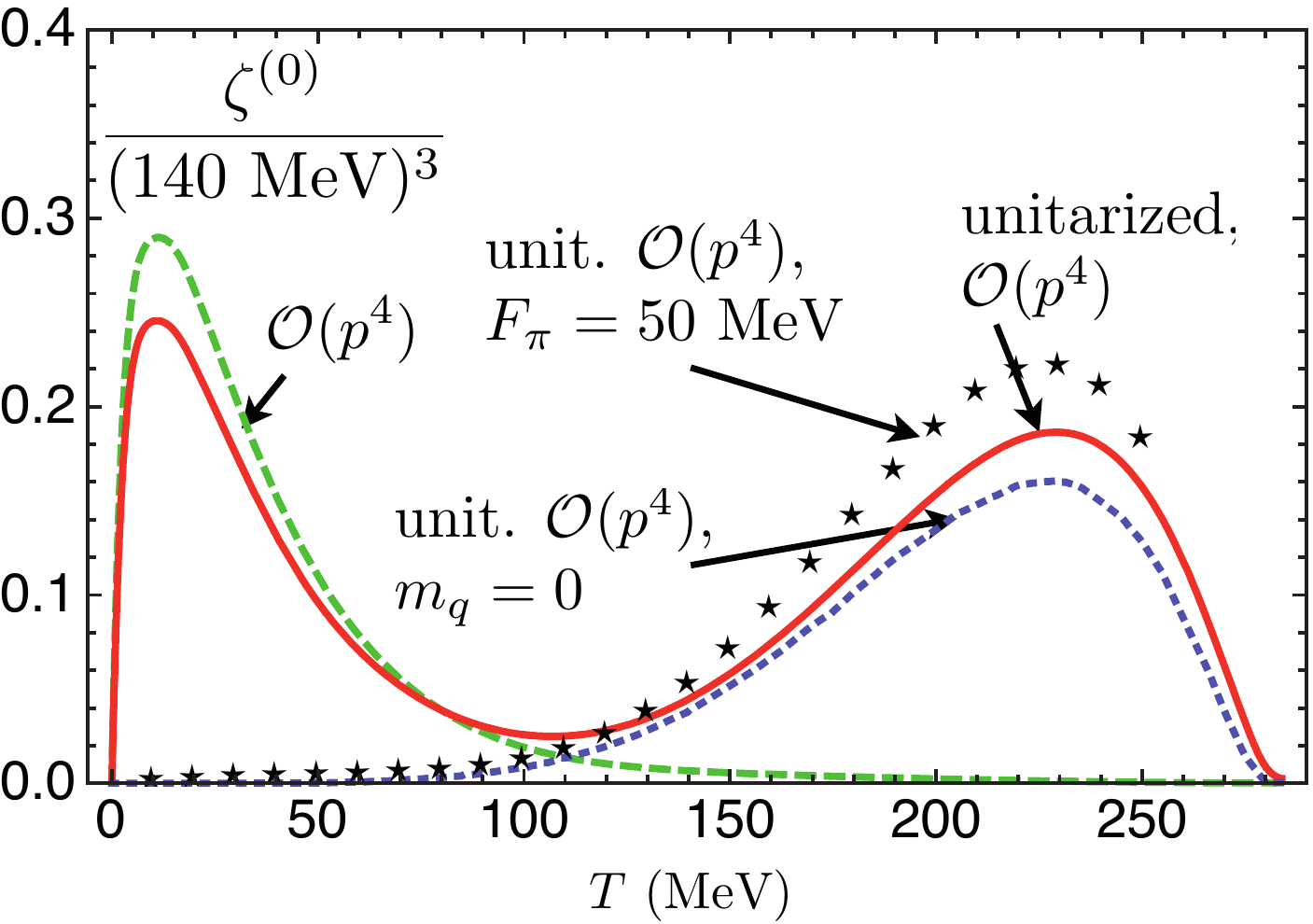}}}
\caption{Lowest-order contribution for the bulk viscosity of a pion gas considering different approximations in the scattering amplitudes.} \label{plotbulk}
\end{figure}

\begin{figure}[h!]
\centerline{\resizebox{0.40\textwidth}{!}{\includegraphics{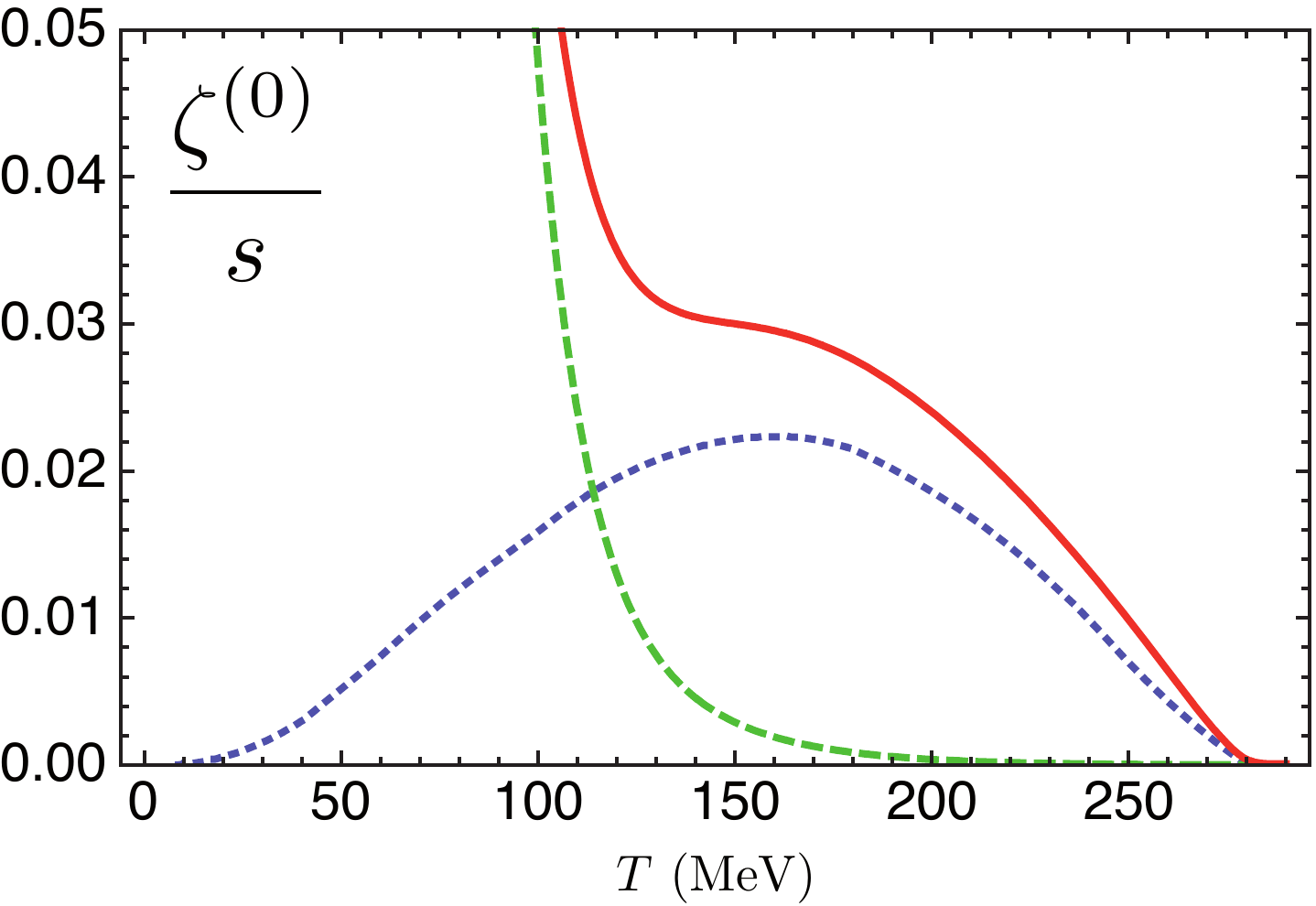}}}
\caption{Lowest order contribution for the bulk viscosity over the entropy density of a pion gas. The curves correspond to the same cases as in Fig. \ref{plotbulk}.} \label{plotzetas}
\end{figure}

Finally, in Figs. \ref{plotbulk} and \ref{plotzetas} we plot the results in \cite{ffgn08} of the lowest-order contribution to the bulk viscosity and the bulk viscosity over the entropy density respectively. We explicitly show the importance of introducing unitarized scattering amplitudes (resonances) in order to reproduce the peak near the phase transition. Nuclear density effects do not change significantly the height of the anomalous peak, as expected from the previous analysis of the conformal anomaly. Comparing with Fig. \ref{plotetas},  our result for the ratio $\zeta^{(0)}/s$ is still smaller than $\eta^{(0)}/s$ near the transition, although the correlation with the conformal anomaly is clear, and that allows to predict larger $\zeta/s$ values if heavier states are included \cite{ffgn08}. Nevertheless, these are extrapolations based on a low-$T$ analysis and must be taken with a grain of salt.

\section{Large-$\text{\it\textbf{N}}_\text{\it\textbf{c}}$ behavior of transport coefficients}

One of the main advantages of our formalism is that we can readily
obtain the parametric dependence with the number of colors $N_c$.
This analysis is interesting, given the theoretical relevance of the
large $N_c$ to describe qualitatively the QCD low-energy sector
\cite{Manohar:1998xv}. In addition, it will confirm some of our
previous qualitative arguments.

 The large-$N_c$  counting of the low-energy constants $\bar
l_i$ can be extracted from that of the $SU(3)$ ones  $L_i$
\cite{Manohar:1998xv,Gasser:1984gg} in the $N_f=2$ limit
\cite{Gasser:1984gg}, while $F_\pi^2=\mathcal{O}(N_c)$. This gives
for the $\pi\pi$ scattering amplitudes $\vert T \vert^2\sim{\mathcal
O}(1/N_c^2)$, regardless  of whether they are unitarized or not, and
therefore, according to (\ref{widthGoity}), we get
$\mathnormal{\Gamma}_p\sim\mathcal{O}(1/N_c^{2})$. This result,
together with the $N_c$ scaling of the thermodynamic quantities
$P\sim\epsilon\sim c_s^2\sim s\sim \mathcal{O}(1)$, which we extract
from \cite{Gerber}, implies that all transport coefficients scale as
$\mathcal{O}(N_c^{2})$ for $M_\pi\neq 0$.

However, in the chiral limit, from the expression of the pressure
\cite{Gerber}

\begin{align}
P=\frac{\pi^2}{30}T^4\left[1+\frac{T^4}{36
F_\pi^4}\ln\frac{\Lambda_p}{T} + {\mathcal O}(T^6)\right]
\end{align}
and taking into account that $\log\Lambda_p\sim \bar l_1 + 4 \bar
l_2\sim\mathcal{O}(N_c)$, we get in this limit $c_s^2-1/3\sim
{\mathcal O}(\ln\Lambda_p/F_\pi^4)\sim{\mathcal O}(1/N_c)$. Now,
since $\zeta/\eta\sim{\mathcal O}[(c_s^2-1/3)^2]$ in the chiral limit
and the previous counting of the width is valid also in this limit,
this means that for $M_\pi=0$, the scaling  of the bulk viscosity is
$\zeta\sim{\mathcal O}(1)$, unlike the other coefficients which
still scale as ${\mathcal O}(N_c^2)$. Summarizing:
\begin{align}
\text{For}\ &M_\pi\neq 0:\notag\\
&\sigma\sim\kappa\sim\zeta\sim\eta\sim\zeta/s\sim\eta/s\sim {\mathcal
O}(N_c^2)\ .\\
\text{For}\ &M_\pi=0:\notag\\
&\sigma\sim\kappa\sim\eta\sim\eta/s\sim {\mathcal O}(N_c^2),\quad
\zeta\sim\zeta/s\sim {\mathcal O}(1)\ .
\end{align}

These scaling relations are consistent with the results we obtained
in the previous sections. The bulk viscosity is suppressed with
respect to the shear viscosity in the chiral limit, as a consequence
of scale invariance, although this is only a parametric dependence
and it does not take into account the anomalous breaking near the
transition \cite{ffgn08}. For $M_\pi\neq 0$ the explicit breaking of
conformal invariance makes the two coefficients comparable, as we
get for very low temperatures, where the mass terms dominate. For
higher $T$, the chiral limit result is again reached asymptotically.
Note also that the $N_c$ scaling for $M_\pi\neq 0$ is compatible
with our leading expressions (\ref{elcondlowt}), (\ref{kappalowt}),
(\ref{etalowt}) and (\ref{zetalowt}). We disagree with the chiral
limit $N_c$-counting for $\zeta$ given in \cite{Chen:07}, where we
believe that the scaling of $\ln\Lambda_p$ discussed above is not
properly accounted for.  The above $N_c$ behavior is also useful in
order to understand the origin of the different conformal-breaking
terms near the transition \cite{ffgn08}. Finally, comparing with
results from high-$T$ QCD is also revealing. From the parametric
expressions given in \cite{Arnold,Dogan} with the scaling
$\alpha_s={\mathcal O}(1/N_c)$, one gets $\eta/s\sim{\mathcal
O}(1)$. This is qualitatively compatible with the idea of $\eta/s$
approaching a minimum when coming from the low-$T$ phase to the
critical region, as we also obtain in our approach. In the high-$T$
regime, $\zeta/\eta$ is also suppressed by an additional
$(c_s^2-1/3)^2$ factor.

\section{Conclusions}

Unitarized Chiral Perturbation Theory provides a consistent framework for the study of transport properties of meson matter. We have shown that, after a suitable modification of the standard ChPT power counting and including unitarity corrections in the scattering amplitudes in order to improve their high energy behavior, one ends up with a reasonable description of transport coefficients for temperatures below the transition. At very low temperatures, our approach meets the predictions of non-relativistic kinetic theory, while at higher $T$ we get an adequate behavior of transport coefficients when compared with existing studies based on the kinetic approach. In addition, we provide phenomenological predictions for the zero-energy photon spectrum and the shear viscosity to entropy ratio which are in fair agreement with data. To obtain these results, we have just considered the dominant diagram, with unitarized scattering in the thermal width for the internal pion lines.

The results obtained within our approach for the bulk viscosity show a clear correlation with the scale anomaly, as suggested by previous works. Our formalism has the advantage of providing a theoretical analysis of transport coefficients for a massive pion gas without relying on lattice results, and therefore it might be useful in order to clarify the relation between the zero-energy limit of spectral functions involving the energy-momentum tensor and the bulk viscosity.

We have also studied the large-$N_c$ limit of the transport coefficients obtained in our approach. The parametric scaling with $N_c$ is consistent with our previous analysis and provides a qualitative description for the behavior of shear and bulk viscosities when approaching the critical region.

Concerning future lines of research, we plan to introduce the effect of the strangeness sector (kaons and eta) which is relevant near the transition, where those states are no longer Boltzmann suppressed. The effect of pion chemical potentials, which has been sketched in our derivation here but not included in the results, is also an interesting extension of our work and will be considered elsewhere \cite{Chemical}.

\section*{Acknowledgements}
We would like to thank D. Kharzeev and R. Venugopalan for very useful conversations. We also acknowledge financial support from the Spanish research projects FPA2004-02602, FPA2005-02327, PR34/07-1856-BSCH, UCM-CAM 910309/08, FPA2007-29115-E and from the F.P.I. Programme (BES-2005-6726).
%
%
%
%
%

\end{document}